\documentclass[12pt,onecolumn,twocolumn]{article}
\usepackage[english]{babel}
\usepackage{times}
\usepackage{mathptm}
\usepackage{supertabular}
\usepackage{dcolumn}
\usepackage{fancyhdr}
\usepackage{rotating}
\usepackage{graphicx}

\renewcommand{\baselinestretch}{1.5}
\begin{document}
\onecolumn
\begin{titlepage}
\date{ }
\title{ Identification of Highly  Deformed Even-Even Nuclides  in the Neutron-
 and Proton-Rich
  Regions of the Nuclear Chart from the 
 $B(E2)\uparrow$ and {$E2$} Predictions in the  Generalized Differential
 Equation Model}
\author{       R. C. Nayak \\
 Department of Physics, Berhampur University, Brahmapur-760007, India. \\
        and  S. Pattnaik \\
 Taratarini  College, Purusottampur, Ganjam, Odisha, India.     }



\maketitle
\begin{abstract}
  We identify  here  possible occurrence of large  deformations in the neutron-
 and proton-rich regions of the nuclear chart from extensive predictions
of the values of the reduced quadrupole transition probability $B(E2)\uparrow$
  for the transition from the ground state to the first $2^+$ state  and
 the  corresponding excitation energy $E2$ of even-even nuclei in the recently
developed  
  Generalized     Differential 
Equation  model  exclusively meant for these physical quantities. 
 This is made  possible  from our analysis of the  predicted  values of
these two physical
quantities and the corresponding   deformation parameters derived from 
them such as the quadrupole deformation $\beta_2$,
the ratio of $\beta_2$ to   the  Weisskopf single-particle  $\beta_{2(sp)}$
and the intrinsic electric quadruplole moment $Q_0$,
calculated  for
 a large number of both known as well as hitherto unknown even-even isotopes 
of   Oxygen to Fermium (Z=8 to 100).
 Our critical analysis of the resulting data
convincingly  support possible existence of large collectivity 
  for  the nuclides 
 $ ^{30,32}Ne, ~^{34}Mg$,   $^{60}Ti$, $^{42,62,64}Cr,~ ^{50,68}Fe$, $^{52,72}Ni$, 
$^{72,70,96}Kr, ~^{74,76}Sr,
 ^{78,80,106,108}Zr$ , $ ^{82,84,110,112}Mo$, $ ^{140}Te, ^{144}Xe$,
 $ ^{148}Ba,~^{122}Ce$, $ ^{128,156}Nd,~ ^{130,132,158,160}Sm$ and
 $ ^{138,162,164,166}Gd$, whose  values of $\beta_2$ are found to exceed 0.3
  and even 0.4 in some cases.
Our findings of large deformations in the exotic neutron-rich regions 
support the existence of another "Island of Inversion"
in the heavy-mass  region possibly caused by breaking of the N=70
 sub-shell closure.

\end{abstract}
\date{ }
\maketitle
\end{titlepage}
\newpage
\setcounter{page}{2}
\section{Introduction}

Studies of the nuclear structure for nuclei lying away from the $\beta$-stable 
valley
of the nuclear chart has been a challenging situation of late, due 
to new phenomena being observed  
  such as  the shell-quenching\cite{que,qu2} of 
the so-called magic shell gaps, and the 
onset of exotic deformations leading to the existence  of the so-called "Island of 
Inversion"\cite{inv,inv2,inv3}. Improved 
experimental  technology  and increased
accuracy of the necessary tools have provided the desired boost
making feasible  for such discoveries. More or less, such issues are   
associated with the onset of increased 
collectivity\cite{inv4,cra}  leading to possible occurrence of large deformations
 of those 
nuclei lying in the exotic regions of the nuclear chart. In this connection
values of the physical quantities such as
 the reduced electric quadrupole transition probability $B(E2)\uparrow$
  for the transition from the ground state to the first $2^+$ state  and
 the  corresponding excitation energy $E2$ of even-even nuclei 
  play 
very decisive   role\cite{rm2} in identifying such  occurrences  of increased 
collectivity.
Particularly  the resulting quadrupole deformation parameters $\beta_2$
and the ratio of $\beta_2$ to the Weisskopf single-particle  
$\beta_{2(sp)}$ derived from them significantly help in this regard.
 Over the years host of such experimental
data for these two physical quantities have led 
Raman et al. \cite{rmn}   to undertake   the well-known Oak-Ridge
Nuclear Data Project \cite{rm1} to make a comprehensive
analysis of all such  data leading to  compilation of 
the desired  adopted data table in the year 1987\cite{rm1} and 2001 \cite{rmn}.
 More recently
Pritychenko et al. \cite{prt} have continued the same Oak-Ridge program
in compiling the newly emerging  data
  for  even-even
nuclei  near  $ N \sim Z \sim $ 28. 

Thus  the study of these two physical quantities $B(E2)\uparrow$ and $E2$ 
 has been under constant investigation 
both by  experimentalists  and  theorists. Several theoretical study of these
quantities have been the epitome of various models and authors [see for instance
Raman et al.'s  \cite{rm2} comprehensive analysis]. Global systematics 
particularly by Grodzins\cite{gro}, Bohr and Mottelson \cite{boh} and Wang et
 al. \cite{wan}
were quite useful in the past. However for local systematics of these 
quantities,  models
in terms of difference equations developed by Ross and Bhaduri \cite{bha} 
  and by Patnaik et al. \cite{pat} were found to be successful to some extent.
 In this regard  our recently developed  
 differential equation model\cite{dem,dm2} for the physical quantity
 $ B(E2)\uparrow$ has been found to be quite successful.
In fact   we could later on succeed in extending \cite{dm3} the same model to include its complementary
physical quantity, namely the excitation energy $E2$. According to  this model
which we may  term it as the Generalized Differential Equation (GDE) model,
the   value of both these quantities for 
 a given even-even nucleus is   expressed in terms of  their derivatives 
with respect to the corresponding neutron and proton numbers N, Z. 
The  same  differential equation 
 in the model has been  further exploited to generate two recursion relations,
 which are mainly
 responsible for the success\cite{dem,dm2,dm3} of the model not only for fitting the 
 known  data,  but also for predicting the unknown   when compared
 with 
the recently compiled  experimental data of Pritychenko et al. \cite{prt} in the $ N \sim Z \sim $   28 region.
In passing, we may note   that we\cite{dem,dm3} could  visualize such a differential
equation  for these quantities  on the basis of their  close similarity in 
reflecting 
the shell-structure  with  the so-called local energy of the 
Infinite Nuclear Matter (INM) Model\cite{inm,in3,in4,in5} of atomic nuclei 
developed over the years primarily  based on the generalized\cite{ehv} 
Hugenholtz-van Hove theorem\cite{hvh} of many-body theory. It may be of interest 
to note that the form of the differential equation in the GDE model 
as well as for the local energy in the INM model are exactly similar to that
of the generalized\cite{ehv} HVH theorem of many-body theory. We may further
stress  here that
 any relation in the form of a differential equation for any physical
quantity is intrinsically sound  enough to possess the desirable feature of
 good predictive ability. In fact this was found to be true behind the 
success\cite{in5} of the INM model as a mass formula and also with the
 presently considered GDE model\cite{dem,dm2,dm3}.

Here in the present work,  we are particularly interested to focus
possible occurrence of increased collectivity leading to identification of
  exotic deformations  for the nuclides lying mostly in the neutron- and
proton-rich (n-rich and p-rich)  regions of the nuclear chart. This is achieved
 from our analysis of the  widely predicted data  made in our  model for
  the two  physical 
quantities $B(E2)\uparrow$ and $E2$,  and from  the  deformation
 parameters calculated from them. Accordingly we used our model first, in 
predicting their values for most of the even-even isotopes 
 lying in the nuclear chart from  Z=8 to 100 (O to Fm)  confined to
 the
known data-set region of Raman et al. \cite{rmn},  and then to the adjacent
 isotopes for
which such values are not yet experimentally available. Then in the second step, we utilized
 these predicted values  in calculating
the relevant deformation parameters, namely  the quadrupole deformation 
 $\beta_2$, the ratio of 
$\beta_2$ to the Weisskopf single-particle
$\beta_{2(sp)}$, and the intrinsic electric quadrupole moment $Q_0$ following  the usual
 model-dependent formalism, in which  nuclei
are treated as having  uniform charge distributions. These calculations
 provide us  the necessary tools  to analyze our data in a better way 
in identifying possible occurrence of increased collectivity and the resulting
exotic deformations.
 
  In the following section 2,  we first of all discuss our model in brief for
 sake of continuity and  fruitful analysis of the resulting data. Section 3 
deals with the usual details of calculation.
 Subsequently we present our results and discuss them in section 4.
 Finally we highlight our main findings in the concluding section 5.

\section{The Generalized   Differential Equation  Model   for $B(E2)\uparrow$ and $E2$}

General features along with the details of the model has been well described 
elsewhere first\cite{dem} for $B(E2)\uparrow$  and secondly\cite{dm3} for $E2$.
Since our main interest here is to analyze  the model predictions for 
identifying exotic deformations, we simply 
highlight its basic equations and features.
The principal equation of the 
model valid 
  for both  $B(E2)\uparrow$ and the corresponding excitation energy $E2$ is
 given by   
\begin{equation}
  \label{c}
\frac{C(N,Z)}{A}= \frac{1}{2}\Big[(1+\beta)\Big(\frac{\partial C}
{\partial N}\Big)_Z +(1-\beta)\Big(\frac{\partial C}{\partial Z}\Big)_N\Big],
\end{equation}
where N, Z and A  refer to the neutron, proton and mass numbers of the given nucleus. 
$\beta$ is the   usual asymmetry parameter (N-Z)/A of the nucleus.
The variable $C$ represents 
     both the physical quantities  $B(E2)\uparrow$ and $E2$.
As we can see, the relation (\ref{c})   connects both $B(E2)\uparrow$
 and $E2$ 
 of a given nucleus 
to their  partial derivatives with respect to the neutron   and proton 
numbers N and  Z. We may state here for sake of a comprehensive understanding,
that the very basis behind its proposition  goes to a
similar equation being satisfied by  the local energy component of the 
ground-state energy
of a nucleus, specifically simulating its shell and deformation behavior
 in the infinite nuclear matter (INM) model \cite{inm,in3,in4,in5} of atomic
nuclei primarily built on the basis of the generalized\cite{ehv} HVH 
theorem\cite{hvh} of many-body
theory. Even though its proposition for 
  these two physical quantities $B(E2)\uparrow$ and $E2$ 
 has been made on close similarity with the
local energy term of the INM model, it can be treated as a  semi-empirical
equation as it has been found\cite{dem,dm3} to be satisfied by  
  them  by virtue of
their slow variation  with neutron and proton numbers N and Z locally. Hence
	the differential Eq. (\ref{c}) for these two physical quantities may
be better termed as  localized semi-empirical equation like the difference 
equations of Ross and Bhaduri\cite{bha} and Pattnayak et al.\cite{pat}.
  we further  like to highlight the interesting fact that the
 form  of the differential equation (\ref{c}) for these two  physical quantities
     , for the local energy $\eta$ of
 the INM model and  the generalized HVH theorem 
concerning energy per nucleon of the asymmetric nuclear matter are all exactly
 similar in nature. Of course the genesis of the local energy relation in the 
INM  model owes its origin to the generalized HVH theorem, whereas  
formulation of the 
differential equation for the two physical quantities
      $B(E2)\uparrow$ and $E2$ simulating  the local energy $\eta$ obviously
got  the same form.
 At the same time however we should
note that while the HVH theorem is an exact theorem  of the many-body theory,
the differential equation (\ref {c}) for all the   physical quantities concerning
 the finite nucleus can be termed as model-dependent.

  Then using  the usual 
forward and backward definitions pair-wise for both  the    derivatives given
by
\begin{eqnarray}
  \label{der} 
\Bigl({\partial C/
  \partial N}\Bigr)_Z &\simeq&{1\over 2}  \Bigl[ C[N+2,Z]-C
  [N,Z]\Bigr], \nonumber \\ 
\Bigl( {\partial
  C/ \partial Z}\Bigr)_N &\simeq&{1\over 2}  \Bigl[ C
  [N,Z+2]-C[N,Z] \Bigr] ,\\
& & and  \nonumber \\
\Bigl({\partial C/
  \partial N}\Bigr)_Z &\simeq&{1\over 2}  \Bigl[ C[N,Z]-C
  [N-2,Z]\Bigr], \nonumber \\ 
\Bigl( {\partial
  C/ \partial Z}\Bigr)_N &\simeq&{1\over 2}  \Bigl[ C
  [N,Z]-C[N,Z-2] \Bigr] ,
\end{eqnarray}
 the following two recursion 
relations in  C would result 
\begin{eqnarray}
\label{b2f}
C[N,Z] &=& {N \over {A-2}}\: C [N-2,Z] + {Z \over {A-2}}\:C 
[N,Z-2]  ,\\
\label{b2b}
 C [N,Z] &=& {N \over {A+2}}\: C [N+2,Z]+{Z \over {A+2}}\: C
[N,Z+2] .
\end {eqnarray}
  These recursion relations   connecting  values of both $B(E2)\uparrow$ and
 $E2$  of the neighboring even-even nuclei from lower  to higher mass and 
vice-verse,
are primarily responsible in reaching out  from known to the unknown terrain of
the nuclear landscape, and thereby facilitate their   predictions
throughout. 
One may further note  that the choice of either
forward or backward definitions for both the two derivatives occurring
in the Eq. (\ref{c})
facilitate  derivation of the close-knit first order recursion relations 
(\ref{b2f} and \ref{b2b}), each   connecting three immediate
neighboring
even-even nuclei  with neutron, proton and  mass numbers differing at best
by two units   in the nucleon space as shown in  Fig. 1(a),
a fact which is of our primary concern. In contrast, mixed
        definitions, i.e, one forward and the backward for the derivatives
 would lead to second order relations
 connecting nuclei having  mass
 numbers   differing  up to  four units  as can be seen in Fig. 1(b)
 and hence are ignored.

 It is essential to stress here that these recursion 
relations not only connect isotopes of the same element but also different 
neighboring elements having proton numbers  Z, Z-2 and Z+2. Therefore these 
recursion relations should not be interpreted  as  interpolation and 
extrapolation formulas. Moreover one should also note that since these 
relations connect  isotopes of the neighboring elements, they facilitate 
prediction of the hitherto unknown data for the desired isotopes of a given
 element using the existing data of the relevant isotopes of the neighboring 
elements,  even if  its own data for the  neighboring isotopes are either not available 
or scantly available. Even these interconnections connecting the isotopes of the
 neighboring elements  provide possible
means of  bridging sharply changing isotopic variations of these two physical
 quantities    across the isotopes of  a given element.

 	In actual practice, we  use   the  known available data in the neighborhood 
 of a given nucleus 
 in the  two recursion relations (\ref{b2f}) and (\ref{b2b}) separately
 to generate its all possible values for $B(E2)\uparrow$  and $E2$.
Since each of  these
relations can be rearranged  in three different ways  by shifting the three
terms occurring in  them from left to right and vice-verse,
 in principle one can generate up to six alternate
values at best for a given nucleus.  This is  however subject to availability of the corresponding data.
Again  each of them being equally probable, 
the predicted value  is then obtained by the arithmetic  
mean of all those  generated values so obtained. We would like to comment here
that this  method of taking the arithmetic mean of the equally-probable 
generated values  for a
given isotope in a way, achieves some sort of uniqueness in the model 
predictions and at the same time  automatically takes care of all possible
local connections in a given  locality. That is why this  scheme has been 
found to be
 successful\cite{dem,dm2,dm3} in our limited predictions made
earlier for both the physical quantities $B(E2)\uparrow$ and $E2$. 
Thus, our  actual calculation procedure  uses the available experimental  data 
 in predicting values of these two physical quantities
 both for  the    known  as well as for the  hitherto unknown even-even 
nuclides.
  The   predictions made in the  first generation
thus  obtained
  for the unknown, are again used  along with the known data
   in the second step to generate the next generation predictions and so on.
This  procedure is continued  to reach out 
more and more neighboring regions of the nuclear chart.
 However we must mention here,  that  although this scheme in principle can be 
continued  as widely as we please in the nuclear chart, in practice,  it is   
terminated   to  avoid accumulation of errors.  Nevertheless, we find that 
three to four generations are sufficient enough to reach out a large number of 
isotopes on either side of the normal $\beta$-stable valley for our present
study.

\section{ Calculation of the Deformation Parameters from  the $B(E2)\uparrow$
and $E2$ Model Predictions }

Following  the procedure laid down in the previous section, we have carried out
the  prediction scheme  in the   model using  the combined data set
 of both Raman et al. \cite{rmn} and   Pritychenko et al. \cite{prt}
near $N\sim Z \sim  28$  as the input  experimental data. Accordingly
the total number of $B(E2)\uparrow$ input data 
  comprises  altogether 330 even-even nuclides spread over
the  entire nuclear landscape ranging from O to Fm (
 Z=8 to 100), while the same  for $E2$ is 557. Since our main interest  in the
 present study is to 
identify  possible occurrence of exotic regions of deformation in the n- 
and p-rich  regions adjacent to the already known data valley, we have
 confined our  calculations  up to 
three to four  generations of our prediction scheme. As a result,  our  present calculations
 have   yielded 
hitherto unknown $B(E2)\uparrow$ data  of 278 adjacent isotopes and $E2$ values of
 175 isotopes apart from for those of the  known data set.

In the next step, we used these   predicted data for
 calculating  the standard deformation parameters such as the quadrupole 
deformation $\beta_2$ and 
the ratio of  $\beta_2$ to
the Weisskopf single-particle $\beta_{2(sp)}$, termed here as $\beta_r$ for
simplicity. We would like to stress here that the value of the quadrupole
deformation 
 $\beta_2$ more or less reflects the nature of collectivity of a 
given nucleus.   Its
 zero value would mean no deformation at all  while its finite  value 
would otherwise 
indicate increasing  deformations or  collectivity of a given  nucleus. In 
general,
its  value  up to 0.1 more or less reflects   spherical nuclei  while
 that of  in the range 0.1-0.2 usually correspond to  normal deformations.
On the contrary  its value  in the range 
0.3-0.5 has been shown\cite{lis,lis2,hyd,sor,naz} to reflect strong deformations in
nuclei  while its  value of
$\approx$0.55-0.65 has been considered\cite{fal,ler} to indicate  super deformation. Therefore any such value beyond 0.3
for a given nucleus  may be considered as to reflect large deformation. Apart
from $\beta_2$, we would also consider a supplementary quantity namely
$\beta_r$ as referred above .
We may point out  here that the ratio $\beta_r$ has been considered\cite{rmn}
 more significant in reflecting possible occurrence of the collective effects
 in nuclei.

 The expressions for these
quantities can be obtained in a model-dependent formalism, in which 
nuclei are treated as to have
uniform charge distributions out to  distance $R(\theta,\phi)$ and zero 
charge beyond. The defining equation  for the  quadrupole deformation 
parameter $\beta_2$ is as usual 
  given by 
\begin{equation}
R(\theta,\phi)=R_0[1+\beta_2Y^*_{20}(\theta)],
\label{rth}
\end{equation}
where $R_0$ corresponds to the radius of a constant density undistorted nucleus and 
$Y_{20}^* $ is the usual axially-symmetric spherical harmonics.
Then the well-known    relation 
that  has been widely used in the literature\cite{rmn} for computing the
deformation parameter 
$\beta_2$ from  the model-independent physical 
quantity $B(E2)\uparrow$ simply follows as
[see for instance Roy \& Nigam\cite{roy}]
\begin{equation}
  \label{b2}
\beta_2=(4\pi/[3Zr_0^2A^{2/3}])[B(E2)\uparrow/e^2]^{1/2}.
\end{equation}
Here $r_0$ is the usual nuclear radius parameter, the value of which is usually
 taken
for compilation of such data as 1.2 fm and $B(E2)\uparrow$ is in units of 
$e^2b^2$. We would like to make a note here that the above expression for 
$\beta_2$ [notations may vary] has been widely used invariably by most of the groups see for
instance Raman et al \cite{rmn,rm2} for extracting
its value from the  experimental  $B(E2)\uparrow$ data. 

For  calculating  the Weisskopf single-particle  $\beta_{2(sp)}$ value,
 its expression can be derived by substituting 
 the corresponding 
Weisskopf single-particle  $B(E2)\uparrow$ value given by

\begin{equation}
  \label{bes}
B(E2)\uparrow_{sp} =2.97\times 10^{-5} A^{4/3} (e^2b^2)
\end{equation}
in Eq. (\ref{b2}). Then the expression for $\beta_{2(sp)}$ simply follows as 
\begin{equation}
  \label{bs}
\beta_{2(sp)}=(4\pi/[3Zr_0^2])\times \sqrt{0.297},
\end{equation}
which numerically can be simplified as 1.59/Z as has been done by Raman et 
al.\cite{rmn}. 
Thus one can calculate the ratio $\beta_r$ using  Eqs. (\ref{b2} and \ref{bs}).

Apart from these two quantities, we  also calculate  another useful
 physical quantity, namely,  the intrinsic electric quadrupole moment
  $Q_0$ in units of $b$  given by

\begin{equation}
  \label{q}
Q_0=\Biggl[{16\pi\over 5} {B(E2)\uparrow\over e^2}\Biggr]^{1/2}. 
\end{equation}
Thus we see that using these Eqs. (\ref{b2}, \ref{bs} and \ref{q}), all the 
relevant deformation parameters can be calculated from $B(E2)\uparrow$.

Before ending this section it is worth mentioning the fact
 that $\beta_{2(sp)}$ as can be seen from Eq. (\ref{bs}) remains  a fractional
 constant for all
the isotopes of a given element, and hence simply acts as a  constant dividing
factor for the quantity $\beta_r$ for all those isotopes.  Thus the  numerical
values of the deformation parameter $\beta_r$ effectively gets enhanced for 
all those 
isotopes  having large deformations by virtue of their larger
 $\beta_2$ values compared to those lying in the normal $\beta$-stable valley 
for a given element. As a result there cannot exist a definite 
 value for this 
quantity to decide whether a particular isotope has a larger or a smaller 
deformation. Therefore the nature of deformation for a given isotope
can only be ascertained by comparing its $\beta_r$ value with those of  its
already known neighboring isotopes. 

\section{Results and Discussion}
\subsection{Identification of Exotic Deformed Nuclides}

As per the details laid down  above, we have first carried out  the 
predictions of $E2$ and
 $B(E2)\uparrow$ data for the desired isotopes lying both in the known and 
the hitherto unknown regions of the nuclear chart.
 Then using these predicted data we  subsequently calculated the  deformation
 parameters $\beta_2$ and 
$\beta_r$ by using the formulas (\ref{b2},\ref{bs}). Our calculations have 
yielded
$B(E2)\uparrow$ values of altogether 608 nuclides which include the  input
data of 330. Similarly our $E2$ predictions  have yielded 732 nuclides that 
include  input data of 557. Since  our main interest being the
identification of the possible  exotic deformations in the hitherto unknown 
data regions,  we 
present  here   in Table 1 only such  data  that are confined to those regions.
 We also present in the same table 
the calculated values of the  deformation parameters $\beta_2$, $\beta_r$ and
 $Q_0$.

 In general, one can easily identify possible
occurrence of the increasing collectivity and the consequent exotic 
deformations specially from the relatively larger 
 values of $\beta_2$ and   $\beta_r$ from 
Table 1. As stated earlier any value of $\beta_2$ larger than 0.3 more or less
reflects higher deformation and increasing collectivity of the given nucleus. 
Such observations  can be  further supplemented by the  increasing values  of
$\beta_r$. 
 However, for sake of conveying better visual display of such occurrences as 
scrutinized from the tabulated values,
  we   graphically present  values of these two 
deformation parameters  for the isotope series as isolines only for those
elements   in the Figs. 2-7.
Accordingly  the graphs displayed in these  figures correspond to such 
  elements having proton number Z=10, 22, 24, 26, 28, 36, 38, 40, 42,
52, 54, 56, 58, 60, 62, 64, 66 and Z=92. We would like to again stress here that
our choice of these elements purely follows from our primary interest
 of identifying
any possibility of exotic deformation in the exotic n- and p-rich regions of the
nuclear chart. Consequently  our close scrutiny of  Table 1 shows increasing
trends in the values of the deformation parameters for either in the n-rich or
p-rich or both for the isotopes of the stated elements except however  for
Z=66 . For instance the
$\beta_2$ value increases from 0.075 to 0.513 with  increasing neutron number
 from N=32 to 38, while $\beta_r$ values increase from 1.043 to 7.116 
 for the element  Z=22. We have intentionally chosen to include the isoline 
for Z=66 just to highlight how such cases need not be considered due to
 the uninteresting nature of  variation of the 
deformation   parameters in the exotic n- and p-rich regions. For sake of comparative analysis and 
continuity in the graphical presentations,  we have also included  in these
graphs  our predictions in the known-data regions along with the 
adopted experimental  values\cite{rmn,prt} to help us to compare the relative
 values of both
$\beta_2$ and $\beta_r$ in our endeavor for identification of possible exotic
deformations. Inclusion of the adopted experimental values  in these graphs
on the other hand would testify the goodness of the model predictions. 
In fact  one can easily identify hitherto unknown-data  isotopes 
from these graphs having   large 
values of $\beta_2$ ($\geq 0.3$)and relatively larger values of $\beta_r$ 
both in  the n- and p-rich regions. Such relatively large
values of these parameters obviously  signify  possible 
occurrence of  exotic deformations for those isotopes.

Now coming to analyzing the individual cases,  we find   [see Figs. 2 (a) and
 Table 1] the values of the   deformation
parameter   $\beta_2$  as 0.59 and  0.63 respectively  at 
 N=20 and 22 for Ne (Z=10). Despite N=20 being a 
 magic number and N=22 is close to it, both these two n-rich
isotopes  are found   to have such  large values of $\beta_2$. On the other
hand the $\beta_r$ values of these isotopes are respectively 3.73 and 3.99,
which are definitely larger than those of its own known   neighbors as can be 
seen from  Fig. 5 (a).
 Thus  such  increase  is a clear indication of the
possible occurrence of higher
deformations in both $^{30}Ne$ and  $ ^{32}Ne$. 
 Fortunately  this finding of ours is  well-supported by 
the recent experimental observation
of enhanced collectivity for $ ^{30}Ne$
  and the resulting disappearance of N=20 shell-closure 
 by Yanagisawa  et al.\cite{inv4}. 
The authors of this experiment have attributed  such occurrence of strong 
collectivity by breaking of the 
N=20 shell-closure  by the intruder states from the pf-shell and  hence are
in favor of its  inclusion in  the "Island of Inversion" \cite{r28,r29}.
 Even the 
neighboring nuclide $^{34}Mg$ has been also found to 
be highly deformed as its $\beta_2$ value is 0.50 [see Table 1] in agreement
 with the experimental finding by Iwasaki et al. \cite{iwa}. Incidentally 
this nuclide has also the same neutron number N=22 as that of $^{32}Ne$.  

Our close scrutiny [see Fig. 4 (a)] also lead us to find possible 
occurrence of large collectivity for 
$^{60}Ti$ as its $\beta_2$ value is 0.51,  which is almost close to that of
super deformation.
 Its $\beta_r$ value has been found to be 7.11 which is again much 
larger compared to its neighboring  known isotopes [see Fig. 7 (a)].  
Accordingly this n-rich isotope of $Ti$ is most likely  be  heavily  deformed 
despite  its neutron number  38 is very close to the semi-magic number 40
and
 its proton number is also very close to the magic  number 20, thereby
 clearly
supporting the possible manifestation for the occurrence of another "Island 
of Inversion" caused by the intruder states from gd-shell\cite{sor}. 

Similarly such occurrences are also seen in case of 
 $ ^{42,62,64}Cr,~ ^{50,68} Fe$  and 
$ ^{52,72}Ni$ [see Figs. 2(b-d), 5(b-d) for $\beta_2$ and $\beta_r$ 
respectively]. 
The $\beta_2$ values for all these nuclei
  lie    in the range 0.29-0.41  signifying large 
collectivity. We also see that the $\beta_r$ values for all these nuclei
lying in the range 4.96-6.13 are well above the corresponding values
of their neighboring known isotopes. 
 Incidentally  these   predictions of ours
are  again  well-supported by  the recent experimental 
observation of increased quadrupole collectivity in $ ^{64}Cr$  and $ ^{68}Fe$
in a Coulomb-excitation experiment by Crawford et al. \cite{cra}.
 It is further  interesting to find more support from  another experimental
 observation
of strong deformation by Sorlin et al \cite{sor} for the isotopes 
$^{60,62}Cr$. In all these n-rich isotopes including $^{60}Ti$ as stated above
, the  N=40
sub-shell closure  most possibly gets broken due to the intruder orbitals $g_{9/2}$
and $d_{5/2}$ leading to strong collectivity in agreement with 
 the conclusions arrived at by Sorlin et al. \cite{sor}.

 Concerning isotopes of Kr, Sr and  Zr (Z=36, 38 and  40),   we find
the exotic isotopes  $ ^{70,72,96}Kr$, $^{74,76}Sr$ and $ ^{78,80,106,108}Zr$  
 to have   values of $\beta_2$ lying in the range  0.40-0.49 [see Figs. 2(e-f),
4(b)], while those of $\beta_r$ 
lie in the range  10.3-11.6 [see Figs. 5(e-f), 7(b)]. Obviously such
 values of $\beta_2$ for these 
isotopes are  quite large enough to signify high deformations in them. 
It is quite satisfying to note here that our present finding of large 
deformation with a $\beta_2$ value of 0.4 for $^{80}Zr $ in fact has been 
well-corroborated by Lister et al.\cite{lis2} long back experimentally.
One can also see that  the n-rich isotope  $^{102}Sr$ [see Figs. 4(b) and 7(b)]
 can also be treated as
 highly deformed as
its $\beta_2$  and $\beta_r$  values are almost close to the above ranges.
For the neighboring element Mo,
we  also find relatively larger values of $\beta_2$ lying in the range 0.39-0.46
 [see Fig. 3(a)] for  the isotopes 
$ ^{82,84,110,112}Mo$. 
 Whereas their $\beta_r$  values lying in the range 10.3-12.2 are quite large
enough compared to their known neighbors  qualifying them to have large
 deformations [see Fig. 6(a)]. Prediction of such strong collectvity for 
the exotic isotopes  $ ^{108}Zr$ and   $ ^{112}Mo$ may be again connected
to the possible existence of another "Island of Inversion" by breaking of the
N=70 sub-shell closure by the intruder states from hfp- shell. Thus 
 the existence
of two  "Islands of Inversion" already detected experimentally 
with the breaking of
shell-closures at N=20 and N=40, and our present prediction of another one at 
N=70 sub-shell closure  appears to be a general feature of nuclear
dynamics in the exotic n-rich regions of the nuclear chart.

Similarly
for  the isotopes of Te, Xe and Ba (Z=52, 54 and 56), we see
relatively higher than normal  deformations for the nuclides $ ^{140}Te, ^{144}Xe$ and 
$ ^{148}Ba$ as their $\beta_2$ values  range from 0.25 to 0.33 [see Fig. 
3(b-d)]. 
The same feature is well reflected with the wide-ranging 
 values of $\beta_r$  from  7.5 to 11.7 [see Fig. 6(b-d)].  
 We would like to further add  here that our calculation also shows the 
p-rich isotope $^{122}Ba$ to  be well-deformed [see Fig. 3(d) and 6(d)] in
agreement with the experimental findings by  Morikawa et al.\cite{mor}.

Concerning   the isotopes of Ce, Nd, Sm and Gd (Z=58, 60, 62 and 64), we 
find the values of $\beta_2$  to lie  in the range 
   0.36-0.46 [see Figs. 3(e-f), 4(c-d)], thereby 
indicating possible occurrence of exotic deformations for the  isotopes
$^{122}Ce  ,~  ^{128,156}Nd,~ ^{130,132,158,160}Sm$ and $ ^{138,162,164,166}Gd$.
These findings  are once again  well supported by the 
 values of $\beta_r$    lying in the range 14.2-17.5 [see Figs. 6(e-f),
7(c-d)].   As usual these values are larger than
the corresponding values of their respective known neighboring isotopes. 

As mentioned earlier, we have also shown  $\beta_2$ and $\beta_r$ isolines for Z=66 in the Figs.
 4 (e) 
and 7 (e) just to highlight the border cases that we have ignored.
We see that  both the deformation parameters   almost remain unchanged
with increase of neutrons and even show  decreasing trends. This is perhaps
 a clear  indication of no substantive  change in nuclear structure.   Hence
 such variation in the deformation 
parameters for the isolines of  other elements that we have not included in
our present study may not be of much  interest. 

Finally coming to the case of Uranium (Z=92) in the very heavy-mass region as
 shown in the Figs. 4 (f)  and 7 (f)
, we find slight increasing trends in the values of the deformation parameters
with the increasing neutron number from N=146 to 154. We see that  $\beta_2$ value 
increases monotonically  from 0.29 to 0.30 and those of $\beta_r$ from 16.73
to 18.62. Therefore we are of the view that the tendency for higher
 deformation possibly exists, but  without having any dramatic change in the 
nuclear structure.  

Thus, in general the regions  in the nuclear chart corresponding to
 the said isotopes discussed above as well as some in the immediate  
neighborhood could be possible regions of large scale exotic deformations, as
 the values of the quadrupole deformation parameter $\beta_2 $ are  closer to
 and even greater than 0.3. As usual the values of $\beta_r$ are
 are relatively larger than their respective known neighboring isotopes.
 Expectedly  such  behavior is well 
supported by the values of the other physical quantity namely the intrinsic 
electric quadrupole moment $Q_0$, which we have
 plotted for all the  isotope series against the neutron  number N 
 in Figs. 8-10. The increasing value of $Q_0$ for those isotopes as seen
from these figures  clearly corroborate  our findings. 

 Even more importantly, all these findings of exotic deformation  listed above   have been well
 borne out 
 with our predicted values of the other physical quantity, namely, the 
excitation energy $E2$ as can be seen from  Table 1.
 Graphical presentations as shown  in Figs.
 11-13 also bear out the same features more convincingly.  We should remember that unlike the  deformation parameters derived 
from the
values of $B(E2)\uparrow$, values of $E2$ are determined completely
independent of the former. Hence the nature of the isotopic behavior of $E2$ is 
expected not only  independent but at the same time opposite to that  of 
the $B(E2)\uparrow$. This is exactly  the case as it should be with $E2$, as 
we see from the complementary nature of the 
graphs displayed in the Figs. 11-13 in contrast to those of the deformation
 parameters $\beta_2$ and $\beta_r$ .
   We find  that the $E2$ 
values of the concerned isotopes  claimed to have large deformations are almost
increasingly small as they lie on the peripheral portions of the graphs,
 in contrast to
the opposite behavior in case of $B(E2)\uparrow$ and   the deformation 
parameters derived from it such as $\beta_2$ , $\beta_r$ and $Q_0$. Numerically
our predicted  E2 values for almost all the isotopes of
  Ne, Ti, Cr, Fe, Ni and Kr  in the low- and medium-mass regions 
claimed to have large   deformations lie in the range 0.35-1.4 MeV. Even
 experimental E2 values of some of these isotopes also lie in the range 0.7-
1.1 MeV [see for instance Fig. 11 for the isotopes $^{50}Fe$, $^{72}Kr$ and $^{72}Ni$].
Whereas both predicted and experimental E2 values of the claimed isotopes 
having high deformations in the heavy-mass region almost lie in the range 
0.07-0.5 MeV. Thus our predicted E2 values convincingly support possible 
existence of exotic high deformations perceived from the predicted $B(E2)
\uparrow$ values and the deformation parameters calculated from it.

Before ending this section, we just want to highlight  here  regarding the 
nature of
agreement of our model predictions  with the adopted  $B(E2)$ and E2 data, 
which of course has been well-demonstrated while developing\cite{dem,dm3} the
 model.
Here the goodness of agreement is once again borne out from the close agreement
 of the derived quantities $\beta_2, ~\beta_r$ and $Q_0$,  and $E2$ itself
as seen from the Figs. 2-13. It is rather  remarkable to see  the 
 nature
of good agreement of the sharply changing isotopic variations of our model 
predictions with 
those of experiment in almost all the cases as seen from the Figs. 2-13, vindicating our assertion made earlier
about the recursion relations (\ref{b2f} and  \ref{b2b}) that
they  should not be treated as interpolation or
extrapolation formulas. The data of the isotopes of the neighboring elements
play decisive role in this regard as the recursion relations 
connect nuclei having   proton numbers Z, Z-2 and Z+2. 
  Thus such remarkable  agreement with the experimental data
\cite{rmn,prt} 
throughout and particularly the nature of sharply changing  isotopic variations
in most cases bear clear testimony of the goodness of the   GDE model.

\subsection{Comparison with the Latest Experimental Data}

Having identified possible  regions of exotic deformation with our predicted
data, it would be of interest to  
 compare  our predicted values of $E2$ and $B(E2)\uparrow$ 
against  any  new  experimental data if available, which we have not included in our
prediction scheme. This would be highly desirable
 as they would provide the test of   reliability of our predictions and 
establish our model for good.
In this connection, we happened to
 come across  a recent  arxive article by  Pritychenko  et al. \cite{pr2}
of their latest data compilation
for some of the  neighboring nuclides adjacent to the already known 
data set.
Obviously this  new adopted data set at least would  give us a good opportunity
 to test our model 
predictions for some if not for all. From our close scrutiny of our 
predicted data given in Table 1 and those of the latest experimental data \cite{pr2},   we find
 that hitherto unknown data of 77 nuclides in case of $B(E2)\uparrow$  and 65 
nuclides in case of of $E2$ are available for this  comparative analysis.
With this view  we  followed Raman et al's \cite{rmn} prescription  of
 comparing   in terms of the order of agreement of our predicted data with 
 these  new experimental data.  
Accordingly  we have presented  the ratio of the predicted values with those 
of the  newly adopted data for $E2$ and $B(E2)\uparrow$ respectively
  in Figs. 14 and 15.
One can easily see that
 60  out of 65 data points for $E2$ (see Fig. 14)  lie within the box indicating
the percentage  of agreement as  92\%. Such an agreement  can be termed  
excellent as per the yardstick stipulated by Raman et al \cite{rmn}.

  Similarly  in case of $B(E2)\uparrow$ predictions, we see that 62 data
points out of 77 lie  within the box (see Fig. 15) with the resulting  
percentage of agreement  as  81\% . Compared to $E2$ the degree of agreement for
$B(E2)\uparrow$ is somewhat less.
However on close scrutiny we find,  that most of the 14 cases that lie 
 outside the box (see Fig. 15)
have relatively larger experimental uncertainty\cite{pr2} to the tune of 40 to
 109\%.  Just to cite few  
examples , the adopted $B(E2)\uparrow$ value  of
$ ^{148}Gd$ is 0.2279 $(^{+.1144}_{-.0548})$,
while the same for $ ^{124}Cd$ is 0.35$\pm.19$ and that of $^{74}Ni$ is 
0.0642$\pm.0442$.
All these values quoted here are in the usual units of $e^2b^2$.
 Thus such large experimental uncertainty would obviously
 affect the actual experimental value.  Secondly 
 for  some of these  14 cases,   the adopted $B(E2)\uparrow$
 values are
themselves  negligibly small, and accordingly any good agreement in 
 such cases may not be feasible to achieve. Just to cite  few such 
examples  for which 
  the adopted $B(E2)\uparrow$ values being  very small  are 0.0096$\pm.0030$, 
0.00373$\pm.00038$ and 0.060$\pm.020$ in case of 
  $ ^{24}Si,~^{50}Ca$, and  $^{56}Ti$  respectively.
 In view of these two aspects we can very well say, 
that  the quality of  agreement of our model predictions for $B(E2)\uparrow$
 with the newly adopted data  is rather  excellent. Thus, more or less we see 
that our  predictions made in our GDE model both for $E2$ and $B(E2)\uparrow$
 very well stand the test of reliability and thereby support once again 
the goodness of the GDE model.

\section{Concluding Remarks}

  In conclusion,  we would like to say that our main concern in the present work
is to identify
   possible occurrence of  large  deformations for some of the even-even
nuclides lying
in the n- and p-rich regions of the nuclear chart  from   our extensive
 predictions for
 the reduced quadrupole transition probability $B(E2)\uparrow$
 and the  complementary   excitation energy $E2$. We have made these 
predictions using 
  our  recently developed  Generalized     Differential 
Equation  model for these two physical quantities.
 These  predictions include
 the hitherto unknown data for the nuclides lying adjacent to the already
known data-regions of  Raman et al. \cite{rmn} for
 most  of the even-even isotopes of   Oxygen to Fermium (Z=8 to 100).
 For sake of facilitating our desired task, we have also included in our 
calculation  values of the  model-dependent 
   deformation parameters  such as $\beta_2$,
the ratio of $\beta_2$ to   the  Weisskopf single-particle $\beta_{2(sp)}$ and
 and  the intrinsic electric quadrupole moment $Q_0$ using  the predicted values of $B(E2)\uparrow$
and $E2$. In this regard,  our critical analysis of the resulting data
convincingly  support possible existence of large collectivity and the 
consequent exotic deformations  for  the 
nuclides 
 $ ^{30,32}Ne, ~^{34}Mg$,   $^{60}Ti$, $^{42,62,64}Cr,~ ^{50,68}Fe$, $^{52,72}Ni$, 
$^{72,70,96}Kr, ~^{74,76}Sr,
 ^{78,80,106,108}Zr$ , $ ^{82,84,110,112}Mo$, $ ^{140}Te, ^{144}Xe$,
 $ ^{148}Ba,~^{122}Ce$, $ ^{128,156}Nd,~ ^{130,132,158,160}Sm$ and
 $ ^{138,162,164,166}Gd$. The quadrupole deformation parameter $\beta_2$
for all these nuclei mostly exceeds 0.3 and even lies in the range 0.45-0.55
for some of them like $ ^{30,32}Ne,~ ^{34}Mg$,  $^{60}Ti$, $^{62}Cr $,
$^{72,70,96}Kr, ~^{74,76}Sr, ^{106,108}Zr$ and   $ ^{82}Mo$. Such large 
collectivity is well supported by the corresponding relatively smaller
values of the 
supplementary physical quantity, namely, the excitation energy E2. The E2 values
mostly lie  in the range 0.35-1.4 MeV for these nuclei in the low- and 
medium-mass region, while the same in heavy-mass region lie in the range
0.07-0.5 MeV. Even some of the available experimental data  in this regard 
do lie in the range  0.7-1.1 MeV.

Our prediction of strong deformation in case of  
 $ ^{30,32}Ne$ and  $^{34}Mg$    in fact are in close agreement with the 
experimental observation by Yanagisawa  et al.\cite{inv4}  
 and  Iwasaki et al. \cite{iwa} respectively 
leading to the existence of the 
 "Island of Inversion" caused by  breaking of the N=20 shell-closure 
by the intruder states from the pf-shell\cite{r28,r29}. Similar predictions in
case of $ ^{60}Ti$, $^{62,64}Cr,~ ^{68}Fe$ also agree with the experimental
findings\cite{cra,sor} again leading to the existence of another "Island of
Inversion" caused by the breaking of the N=40 sub-shell closure by the 
intruder states from the gd-shell. Thus such agreement with the experimental
findings in the medium-low and medium mass nuclei in the exotic n-rich regions
  have made us to conjecture the existence of another "island of Inversion" 
in the heavy-mass region possibly caused by breaking of the N=70 sub-shell
closure by the intruder states from the hfp-shell as we find strong deformation
for the nuclides  $ ^{108}Zr$ and  $ ^{112}Mo$. Thus it appears that the
existence of 
such "Islands of Inversion" in the exotic n-rich regions of the nuclear chart
may be a general feature of nuclear dynamics waiting for to be explored by 
future experiments. In fact analysis\cite{in5} of two
 two-neutron
 separation energy systematics derived from mass predictions in the INM model
of atomic nuclei supports the existence of such islands   in the  heavy-mass
 n-rich region of the nuclear chart apart from the ones
   in the lower- and  medium-mass regions.

Apart from serving the primary purpose of the present work in predicting
exotic deformations in the exotic regions of the nuclear chart as highlighted 
 above, we also observe 
rather good agreement of our predictions with the adopted experimental data.
Even our model  could reproduce  the sharply changing isotopic variations of 
  the two physical quantities $B(E2)\uparrow$ and E2 in agreement  with 
those of experiment, vindicating our assertion 
that  the recursion relations (\ref{b2f} , \ref{b2b}) derived in the model
  should not be treated as interpolation or
extrapolation formulas. In this regard the interconnecting relations 
connecting the neighboring elements having proton
 number Z, Z-2 and Z+2 facilitate achieving this.
This 
supplements  our earlier observation of good agreement with experiment while developing\cite{dem,dm3} the model.

Even to our satisfaction,  we could further succeed in establishing the goodness
of the model
in comparing some of our predictions with the  latest  experimental
 data \cite{pr2} which we have not included in our
prediction process. In this respect
 it is quite remarkable  to find,  that the
quality   of agreement of our predictions for both these two physical 
quantities $B(E2)\uparrow$ and $E2$ is rather excellent.

\centerline {\bf	ACKNOWLEDGMENTS}

 One of us (RCN)   acknowledges some useful discussion with  R. Sahu of 
the  Department of Physics (BU) 
  regarding the range of possible values of the quadrupole deformation
 parameter for relevant deformations in  nuclei.

\~~~~~~
\newpage
\begin{figure}[bth]
\includegraphics[width=5.in, height=5in,angle=-0]{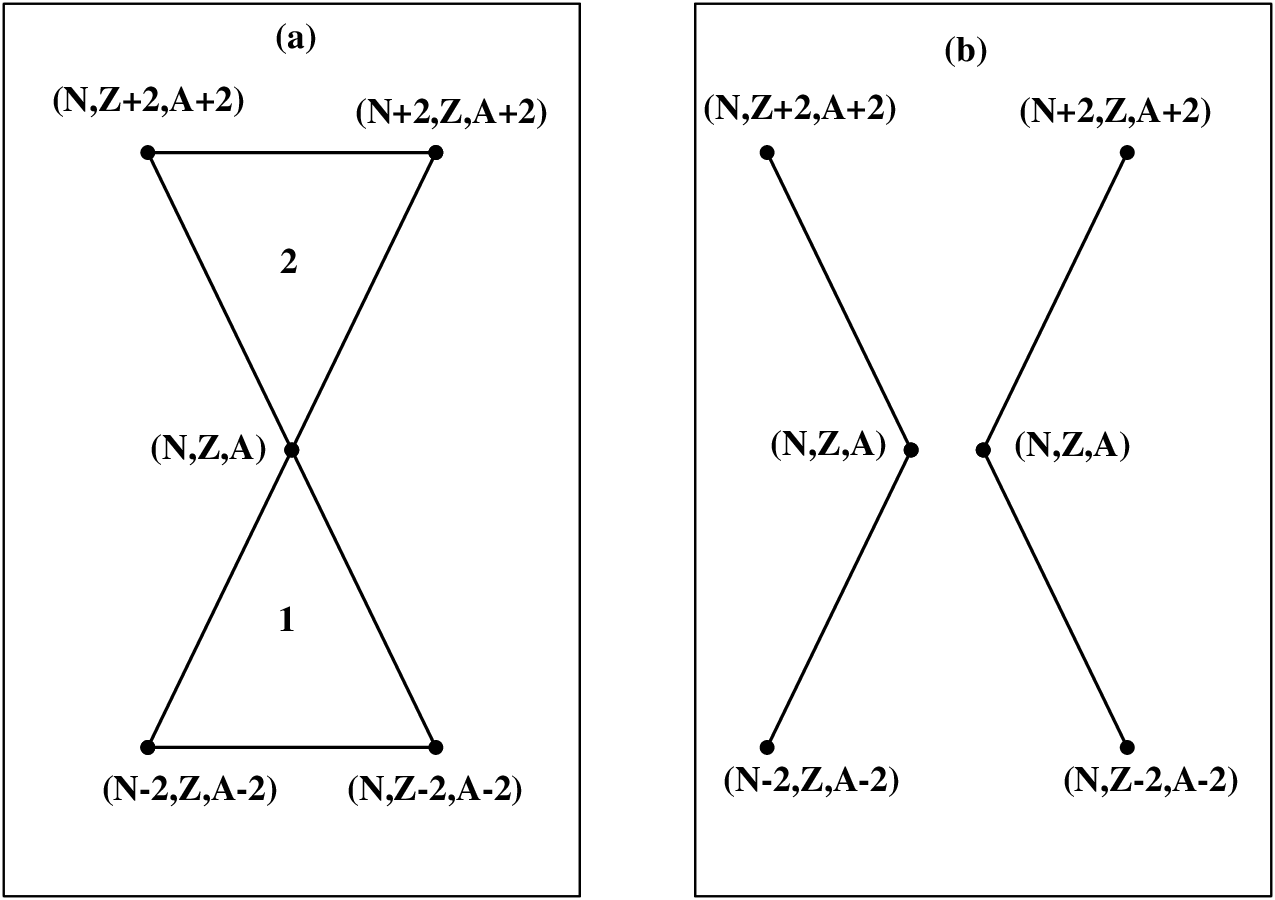}
\renewcommand{\baselinestretch}{1.}
\caption{Schematic diagram showing how  the   recursion relations connect
the neighboring even-even nuclei. (a) corresponds to the first order relations
 (\ref{b2f},\ref{b2b}) connecting nuclei shown here as the vertices of the
two triangles 1 and  2, while (b) shows those of the  second order relations
(see text for details).}
\end{figure}

\newpage
\begin{figure}
\includegraphics[width=7.0in, height=7.0in,angle=-0]{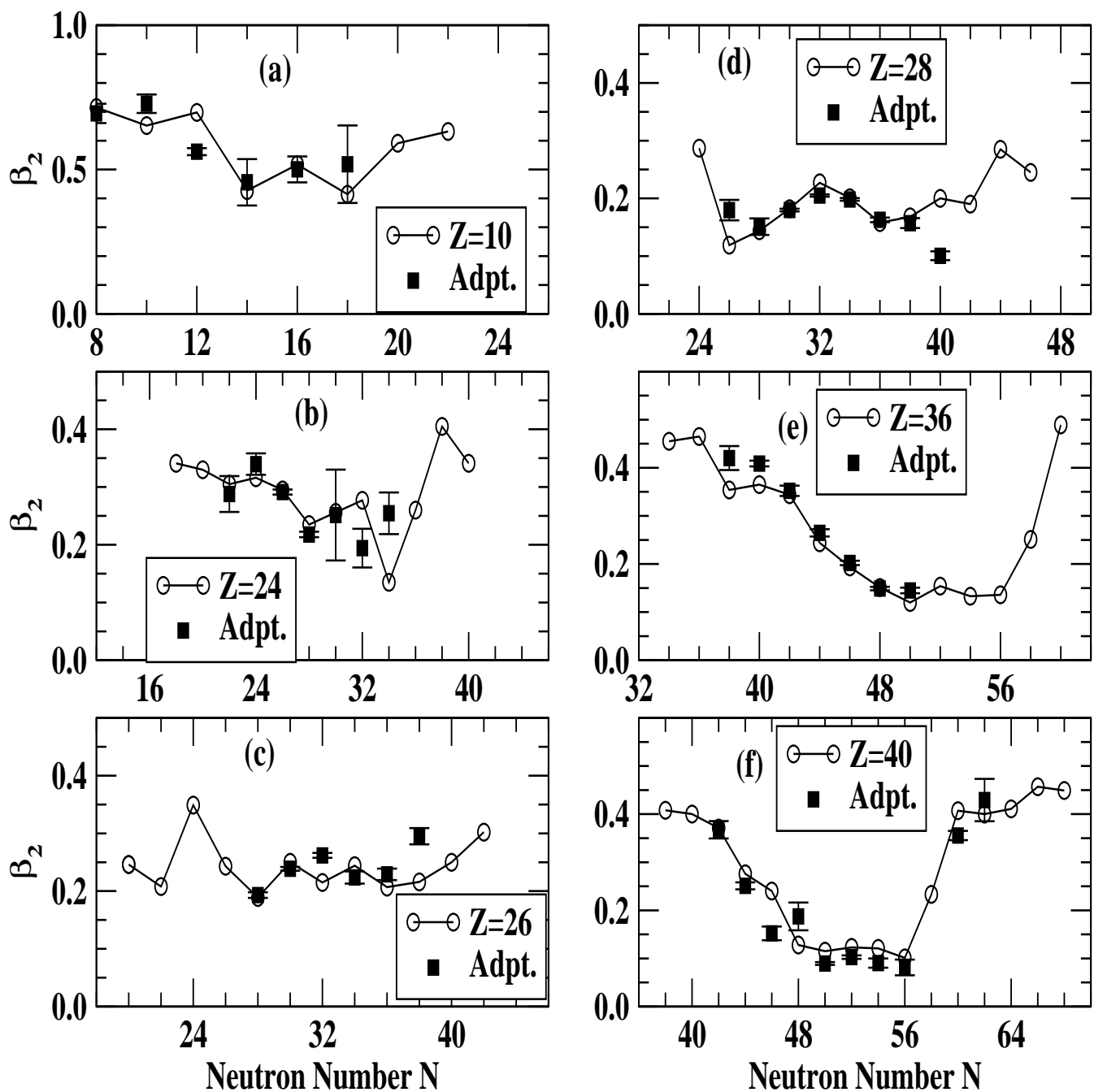}
\caption{Values of the calculated quadrupole deformation parameter $\beta_2$
 (see text)  for the
 isotope series  Z= 10, 24, 26, 28, 36 and 40  plotted here as isolines against
 neutron number N. Thick
lines connecting open circles  represent  our predicted values while solid 
squares with vertical lines   marked as [Adpt.]
 correspond to those of the adopted values\cite{rmn,prt}.
The vertical lines as usual represent the uncertainty in the adopted values.}
\label{beta1}
\end{figure}
\newpage

\begin{figure}
\includegraphics[width=7.0in, height=7.0in,angle=-0]{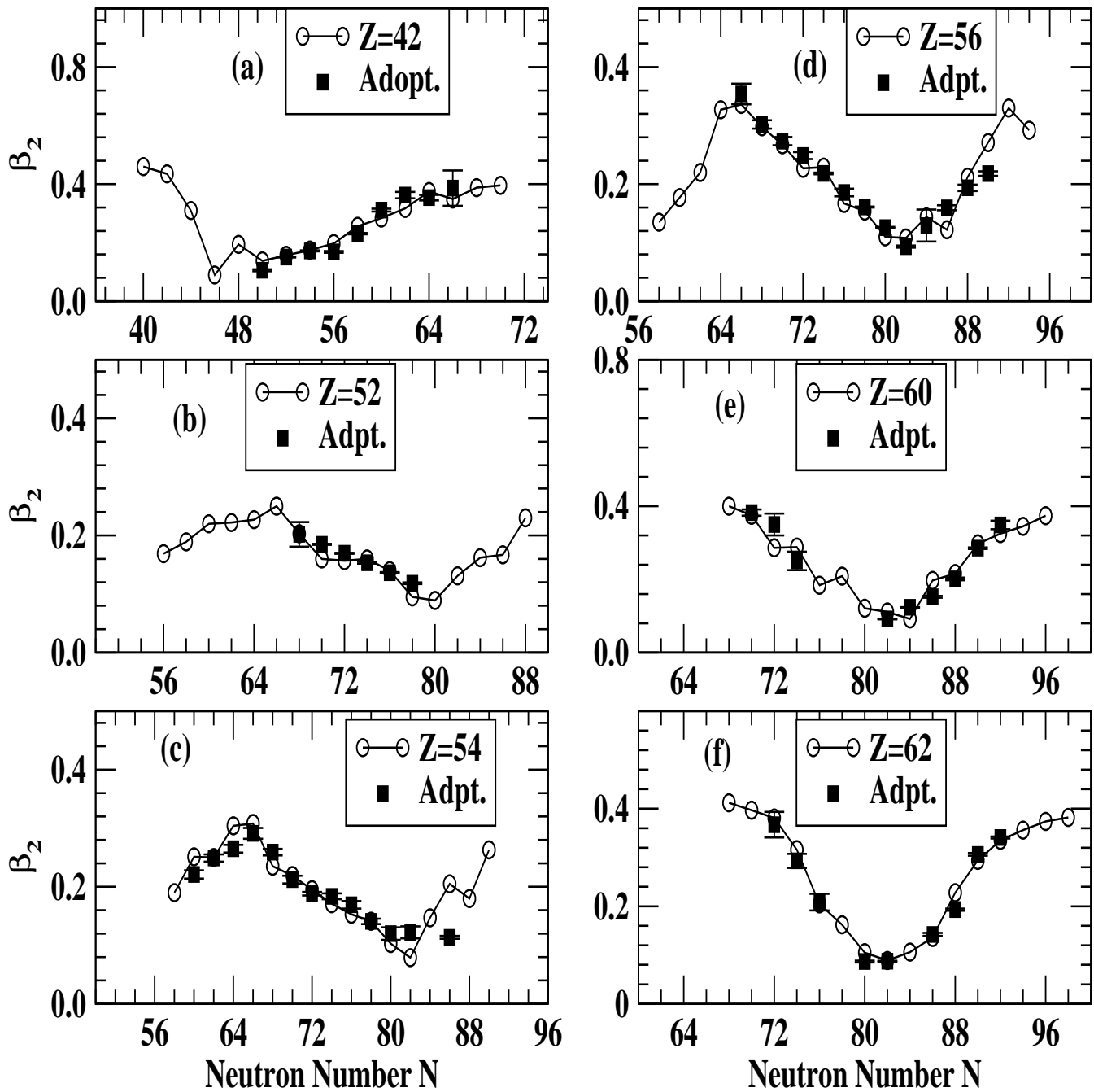}
\caption{Same as Fig. \ref{beta1} but for Z=42, 52, 54, 56, 60  and 62.}
\label{beta2}
\end{figure}
\newpage

\begin{figure}
\includegraphics[width=7.0in, height=7.0in,angle=-0]{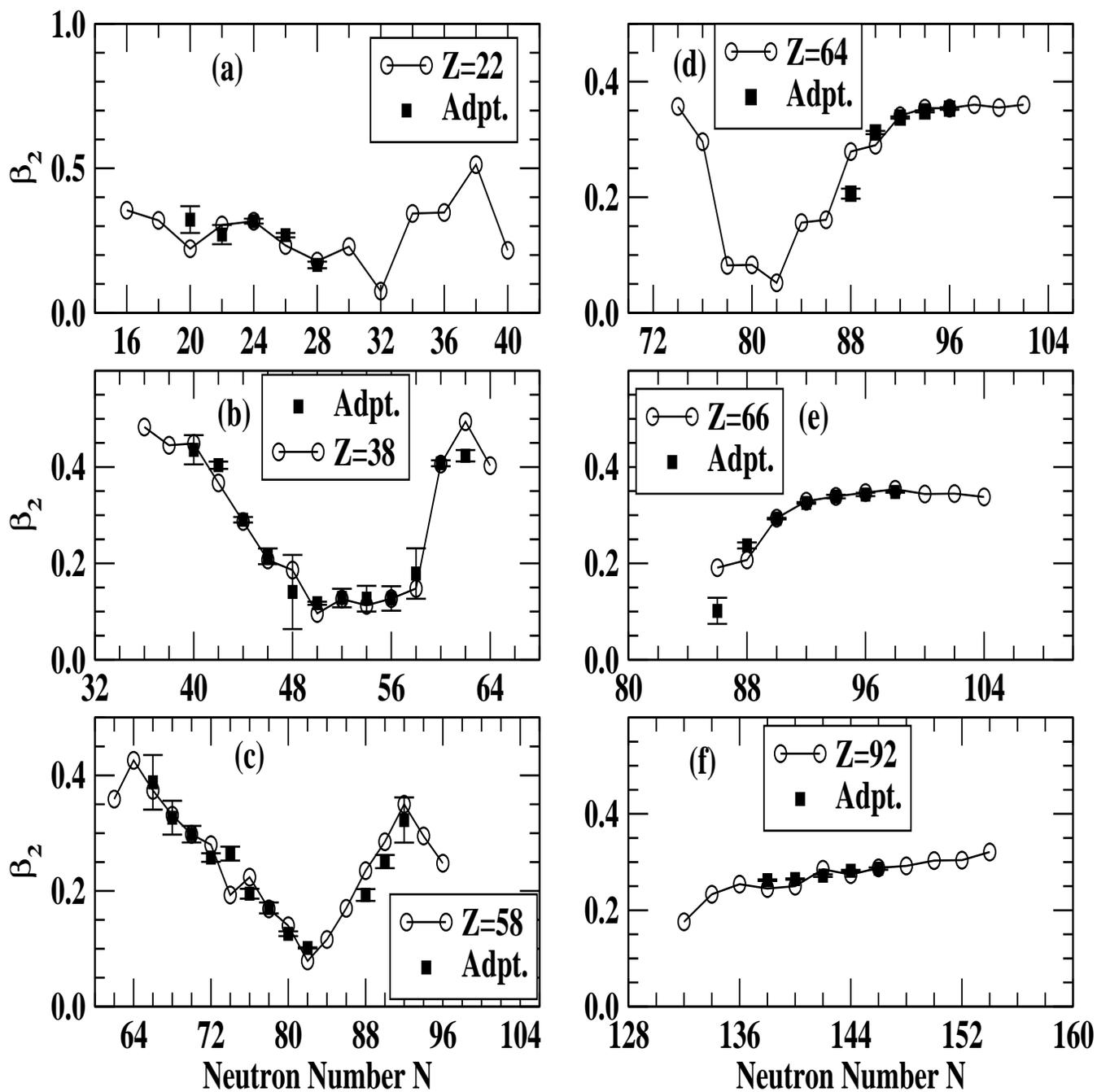}
\caption{Same as Fig. \ref{beta1} but for Z=22, 38, 58,  64, 66  and 92.}
\label{beta3}
\end{figure}

\newpage
\begin{figure}
\includegraphics[width=7.0in, height=7.0in,angle=-0]{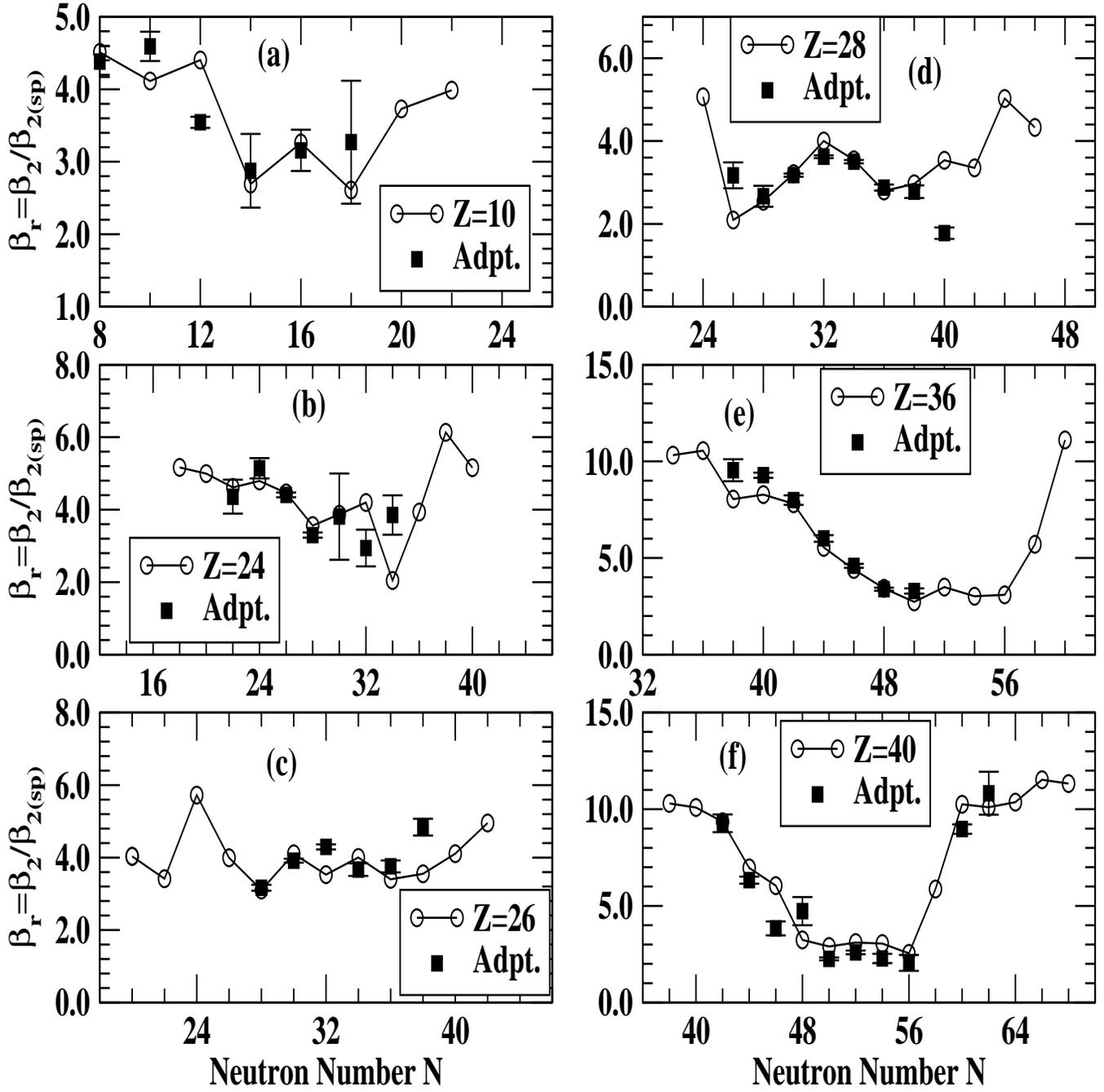}
\caption{Similar to Fig. \ref{beta1}  but for the values of the deformation
parameter $\beta_r$  (see text)  for the  series Z=10,24, 26, 28,36 and 40.}
\label{bet1}
\end{figure}

\newpage
\begin{figure}
\includegraphics[width=7.0in, height=7.0in,angle=-0]{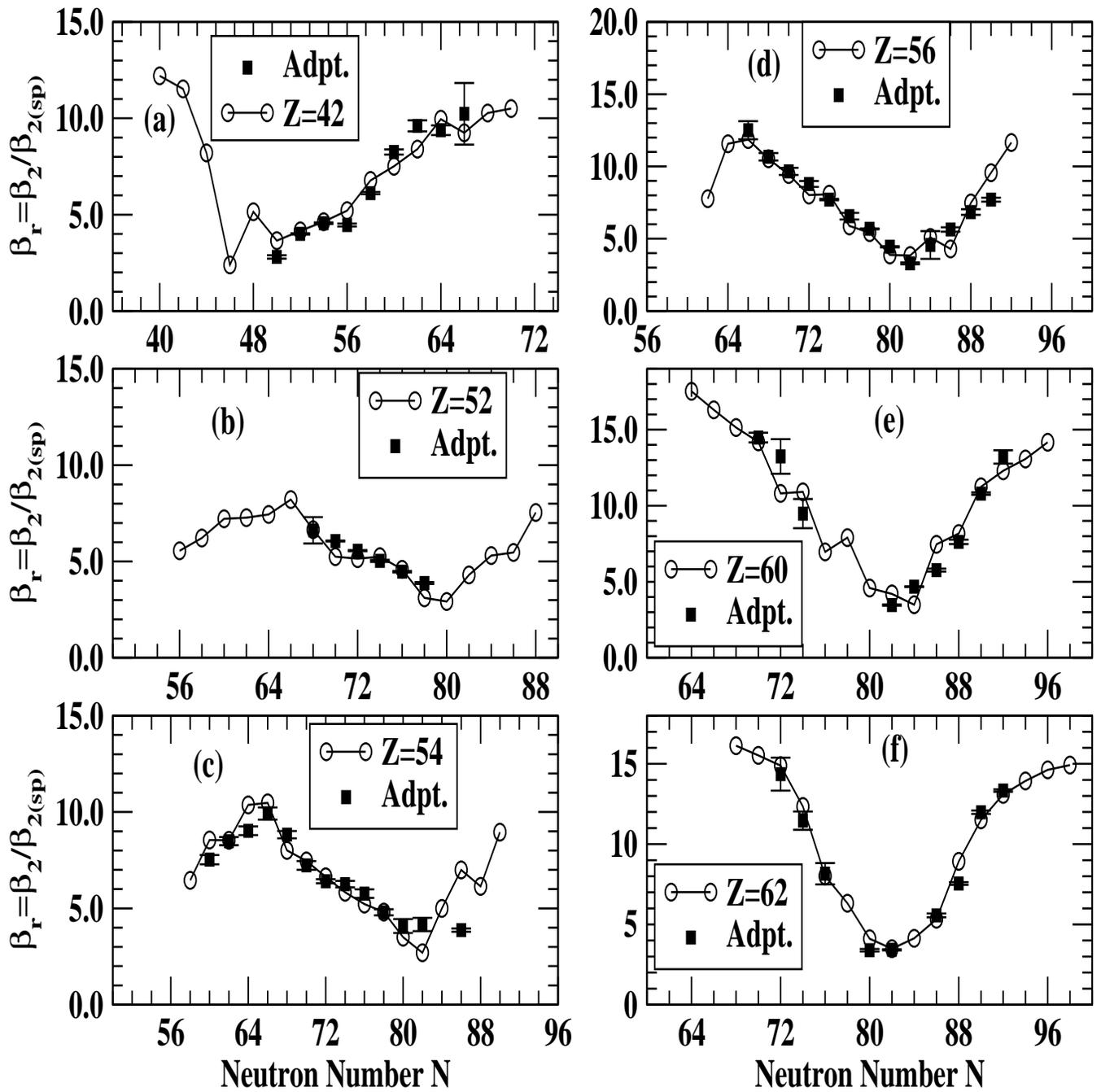}
\caption{Similar to Fig. \ref{beta2} but for the values of the deformation
 parameter $\beta_r$  for Z=42, 52, 54, 56, 60  and 62.}
\label{bet2}
\end{figure}

\newpage
\begin{figure}
\includegraphics[width=7.0in, height=7.0in,angle=-0]{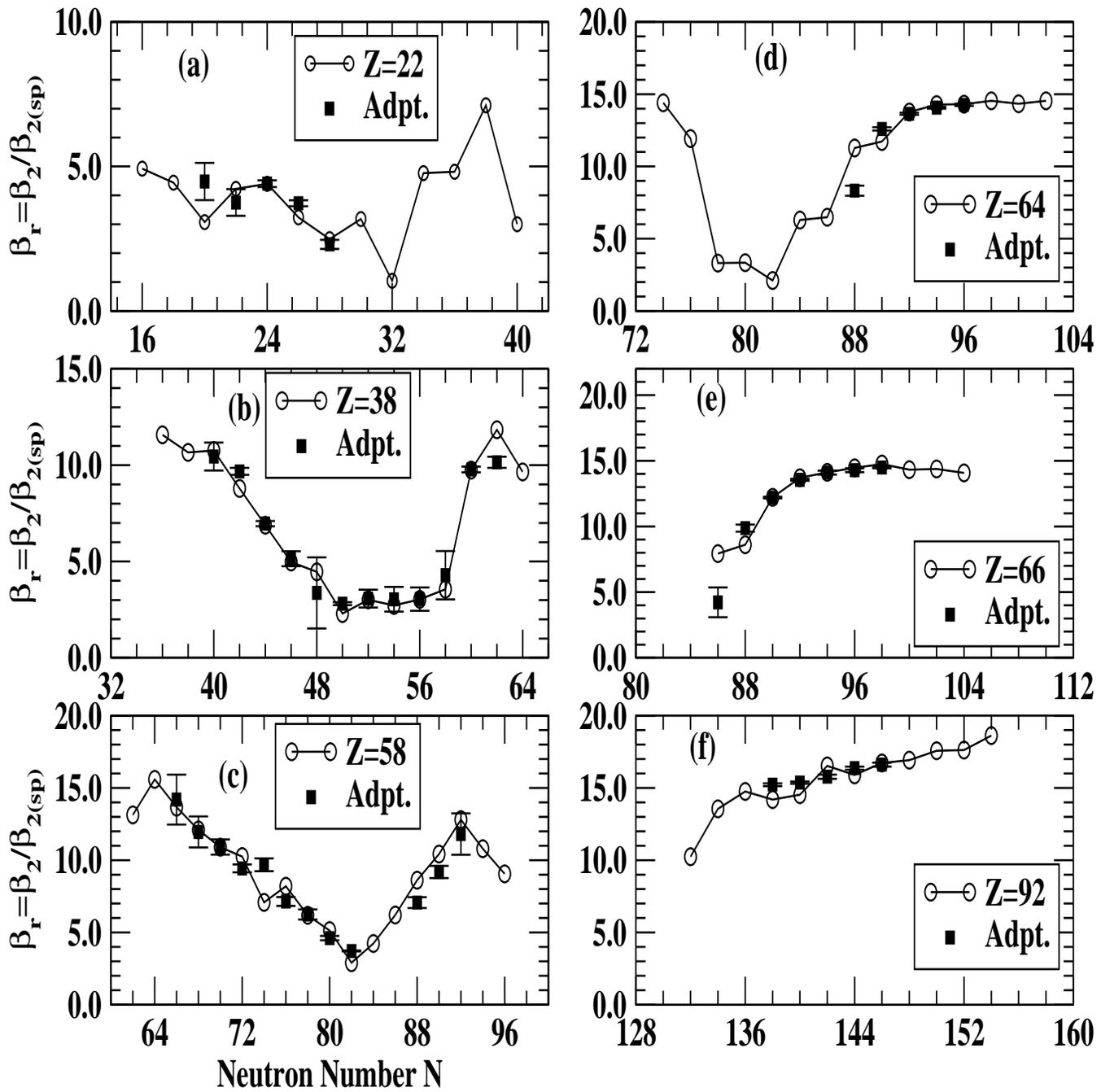}
\caption{Similar to Fig. \ref{beta3}  but for the values of the deformation
 parameter $\beta_r$  for Z=22, 38, 58, 64, 66  and 92.}
\label{bet3}
\end{figure}

\newpage
\begin{figure}
\includegraphics[width=7.0in, height=7.0in,angle=-0]{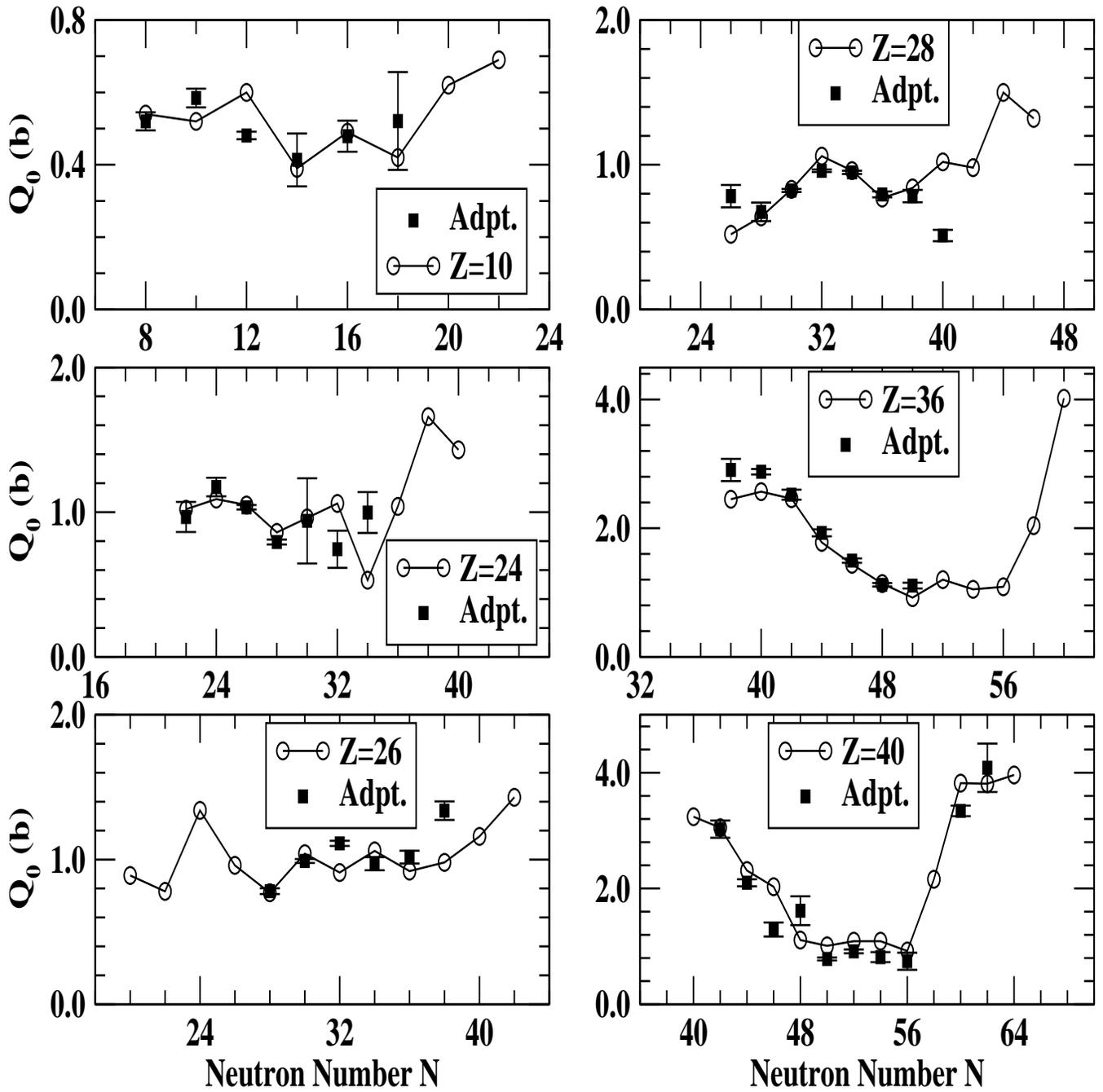}
\caption{Similar to Figs. \ref{beta1}   but for the values of the Intrinsic
Electric Quadrupole Moment
 $Q_0$  for the  series Z=10, 24, 26, 28, 36 and 40}
\label{q01}
\end{figure}

\newpage
\begin{figure}
\includegraphics[width=7.0in, height=7.0in,angle=-0]{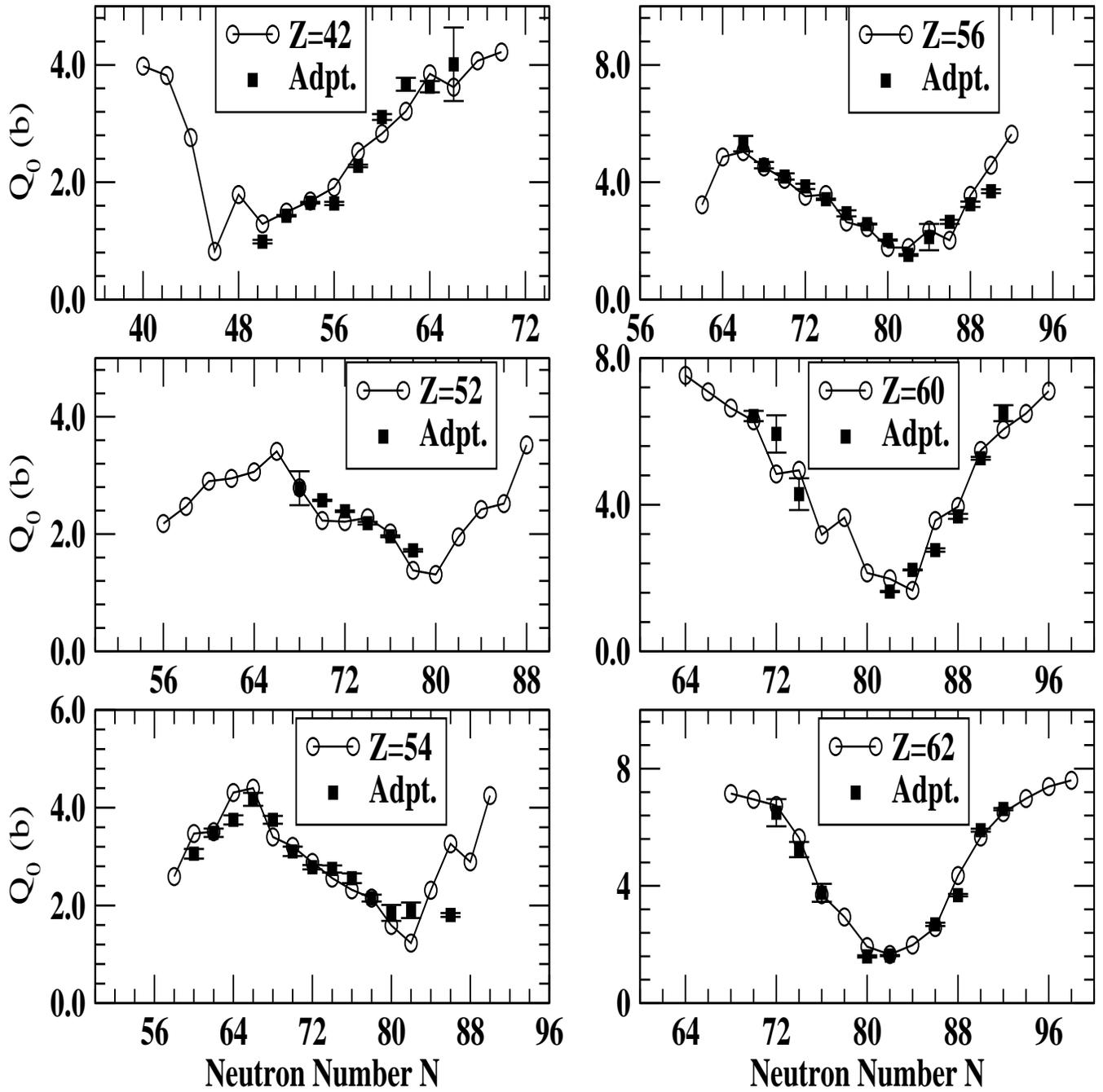}
\caption{Similar to Figs. \ref{beta2}  but for the values of the Intrinsic Electric
Quadrupole Moment  $Q_0$  for the  series Z=42, 52,
54,56,60 and 62 }
\label{q02}
\end{figure}
\newpage
\begin{figure}
\includegraphics[width=7.0in, height=7.0in,angle=-0]{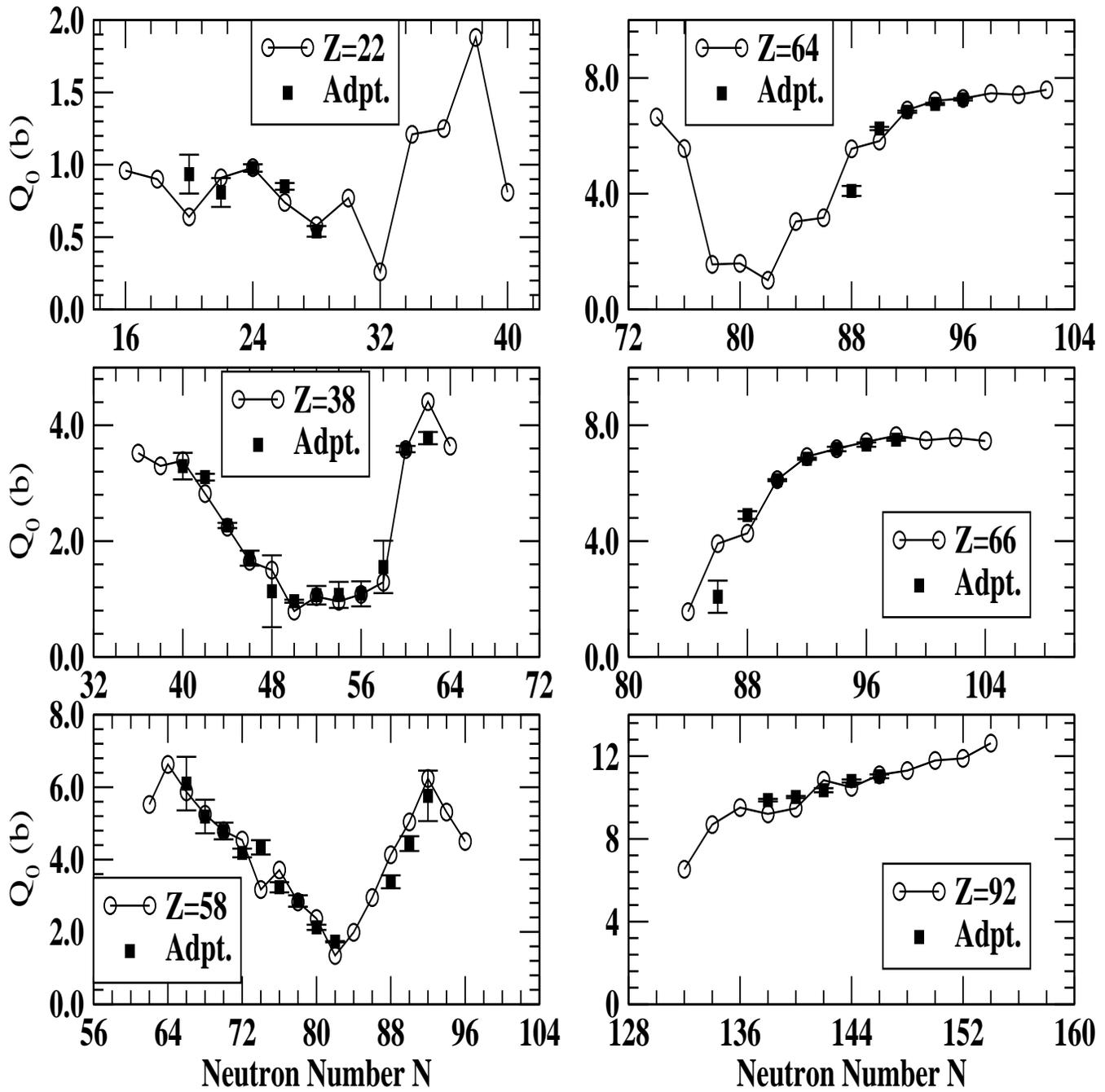}
\caption{Similar to Figs. \ref{beta3}  but for the values of the Intrinsic Electric
Quadrupole Moment  $Q_0$  for the  series Z= 22, 38, 58, 64, 66  and 92.}
\label{q03}
\end{figure}

\newpage
\begin{figure}
\includegraphics[width=7.0in, height=7.0in,angle=-0]{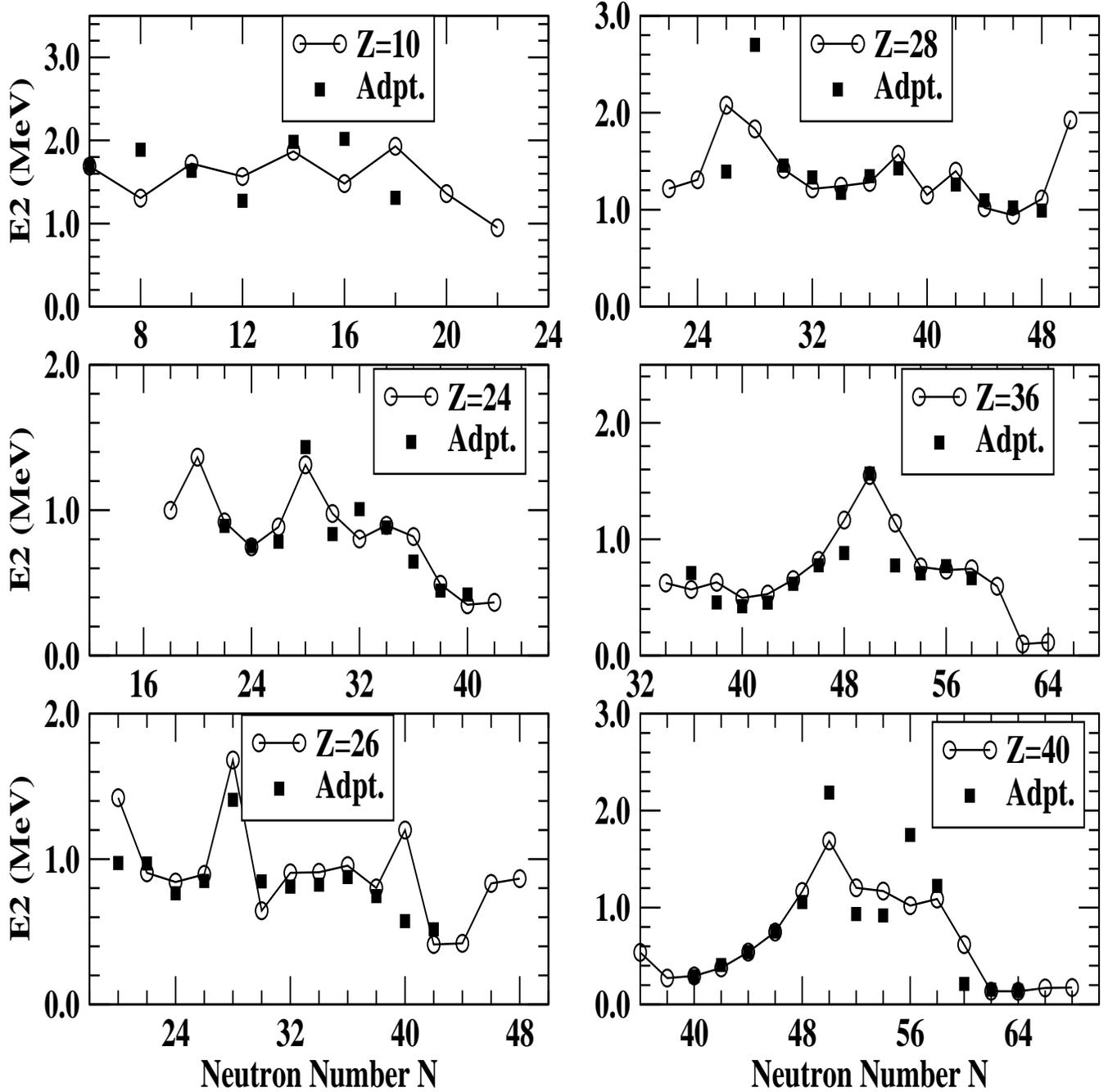}
\caption{Similar to Fig. \ref{beta1}  but for the values of the excitation
energy E2 (see text)   for the  series Z=10,24, 26, 28,36 and 40. However
adopted data are shown without uncertainties as these values are  very
small.}
\label{e1}
\end{figure}

\newpage
\begin{figure}
\includegraphics[width=7.0in, height=7.0in,angle=-0]{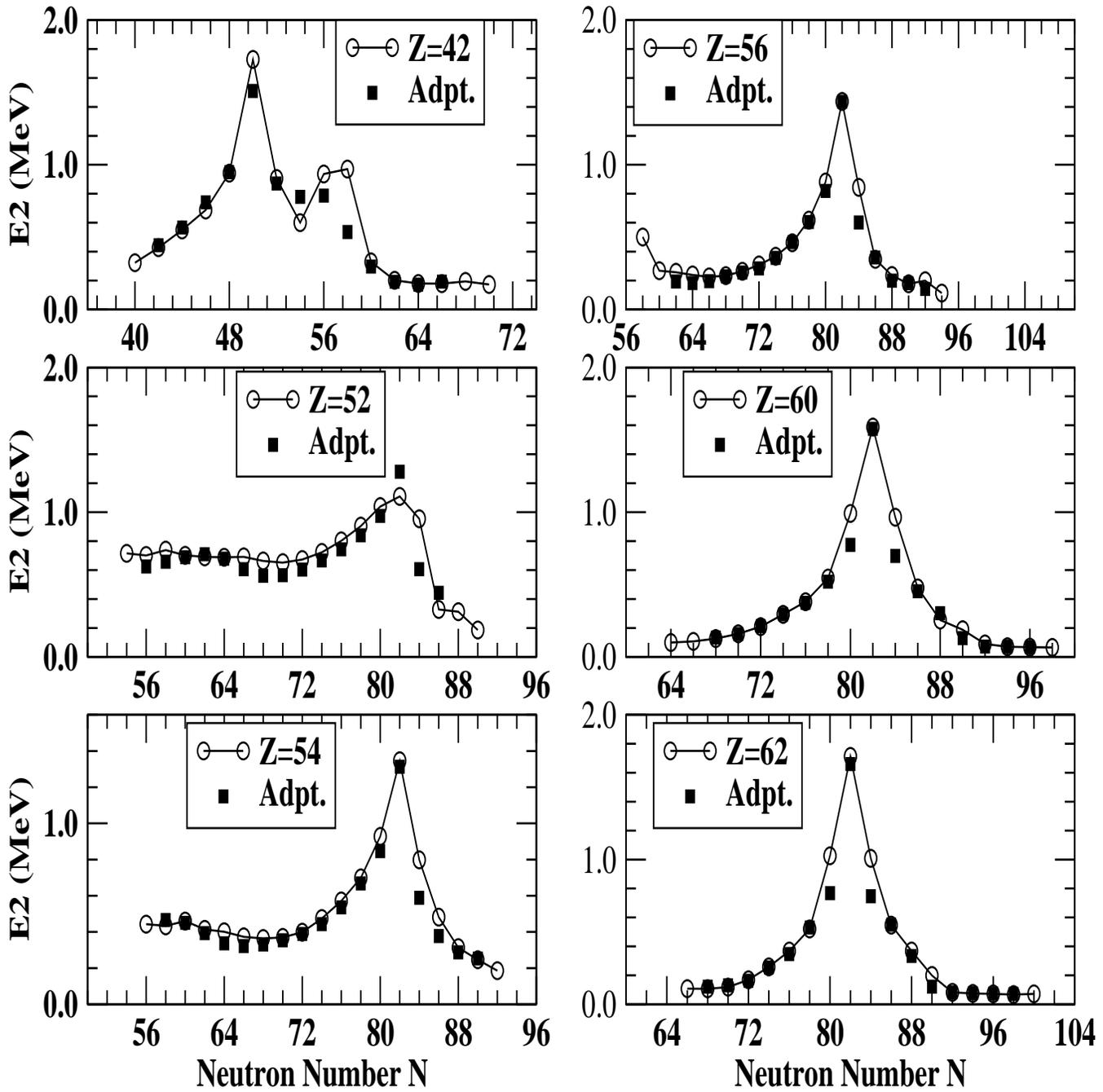}
\caption{Similar to Fig. \ref{e1}  but for the values of the excitation energy
E2 (see text)   for the  series Z=42, 52, 54, 56, 60 and 62.}
\label{e2}
\end{figure}

\newpage

\begin{figure}
\includegraphics[width=7.0in, height=7.0in,angle=-0]{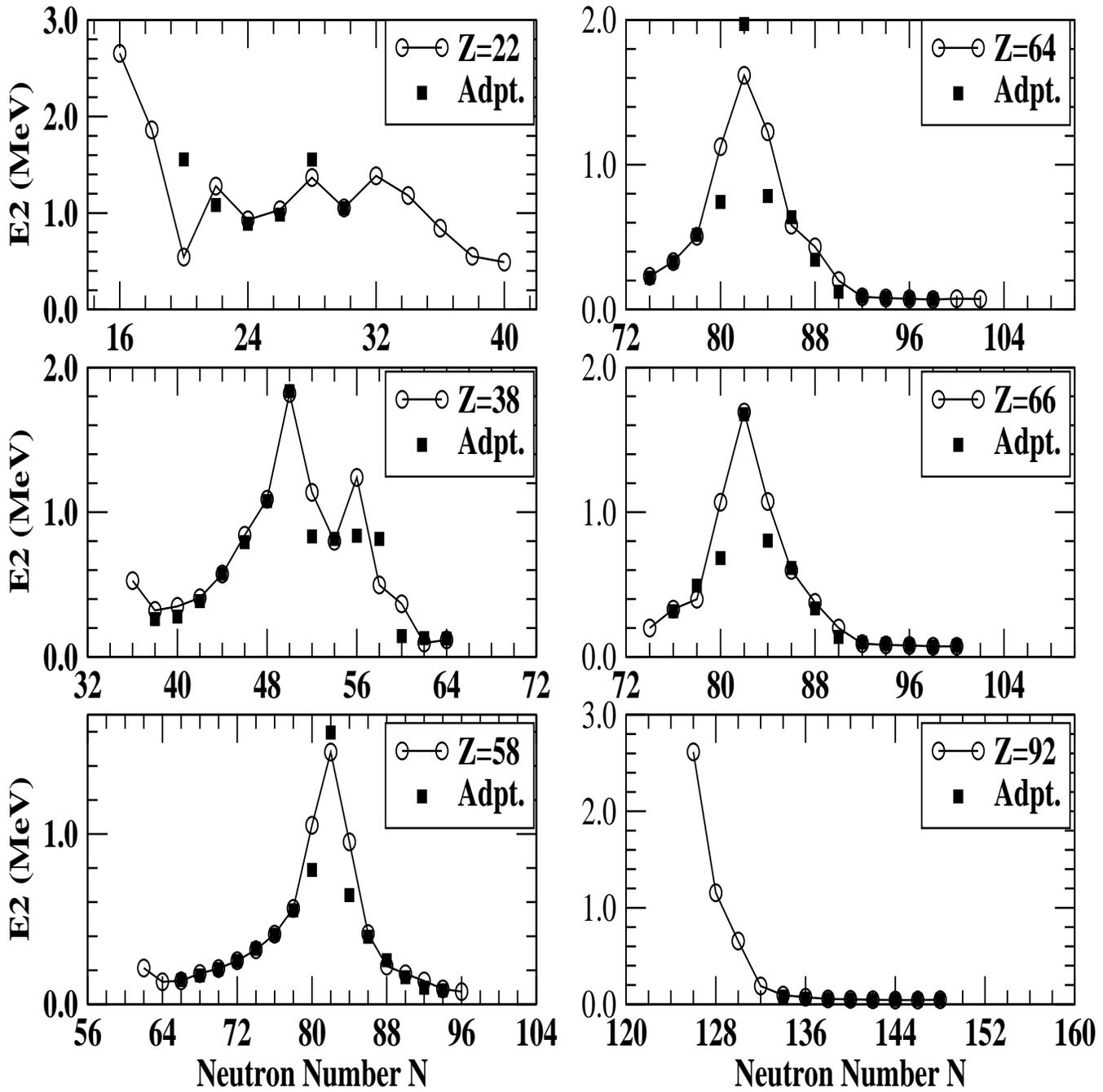}
\caption{Similar to Figs. \ref{e1} and \ref{e2}  but for the values of the excitation energy
E2 (see text)   for the  series Z=22, 38, 58,  64, 66 and 92.}
\label{e3}
\end{figure}

\newpage
\begin{figure}
\includegraphics[width=7.0in, height=7.0in,angle=-0]{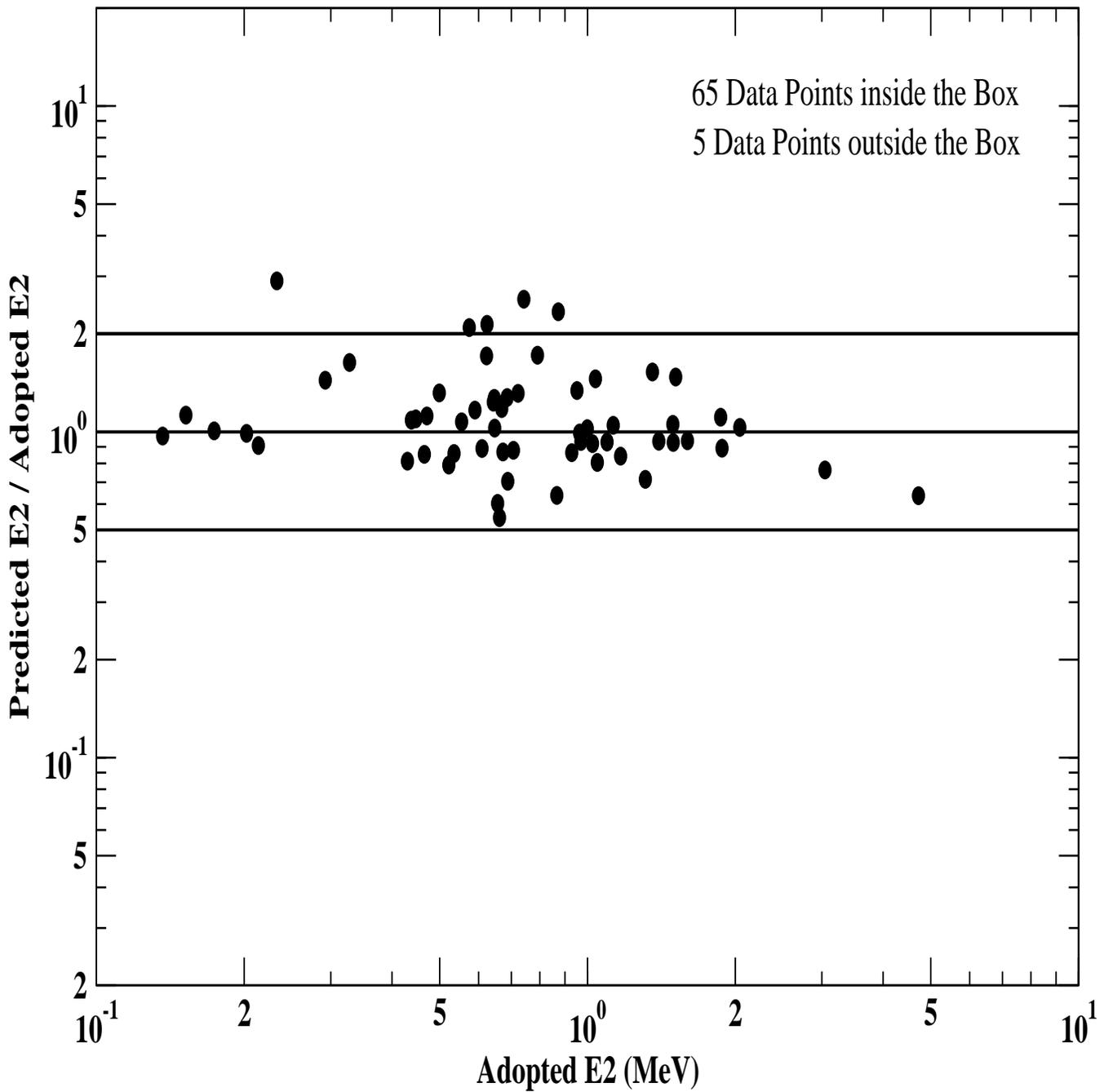}
\caption{Comparison between the  model predictions for the excitation energy
 $E2$ with the latest adopted  experimental data
\cite{pr2} (see text) plotted here in the form of their ratio versus those
of the adopted experimental data. The data points lying inside the box indicate the
degree of agreement within a factor of two. }
\label{e2l}
\end{figure}

\begin{figure}
\includegraphics[width=7.0in, height=7.0in,angle=-0]{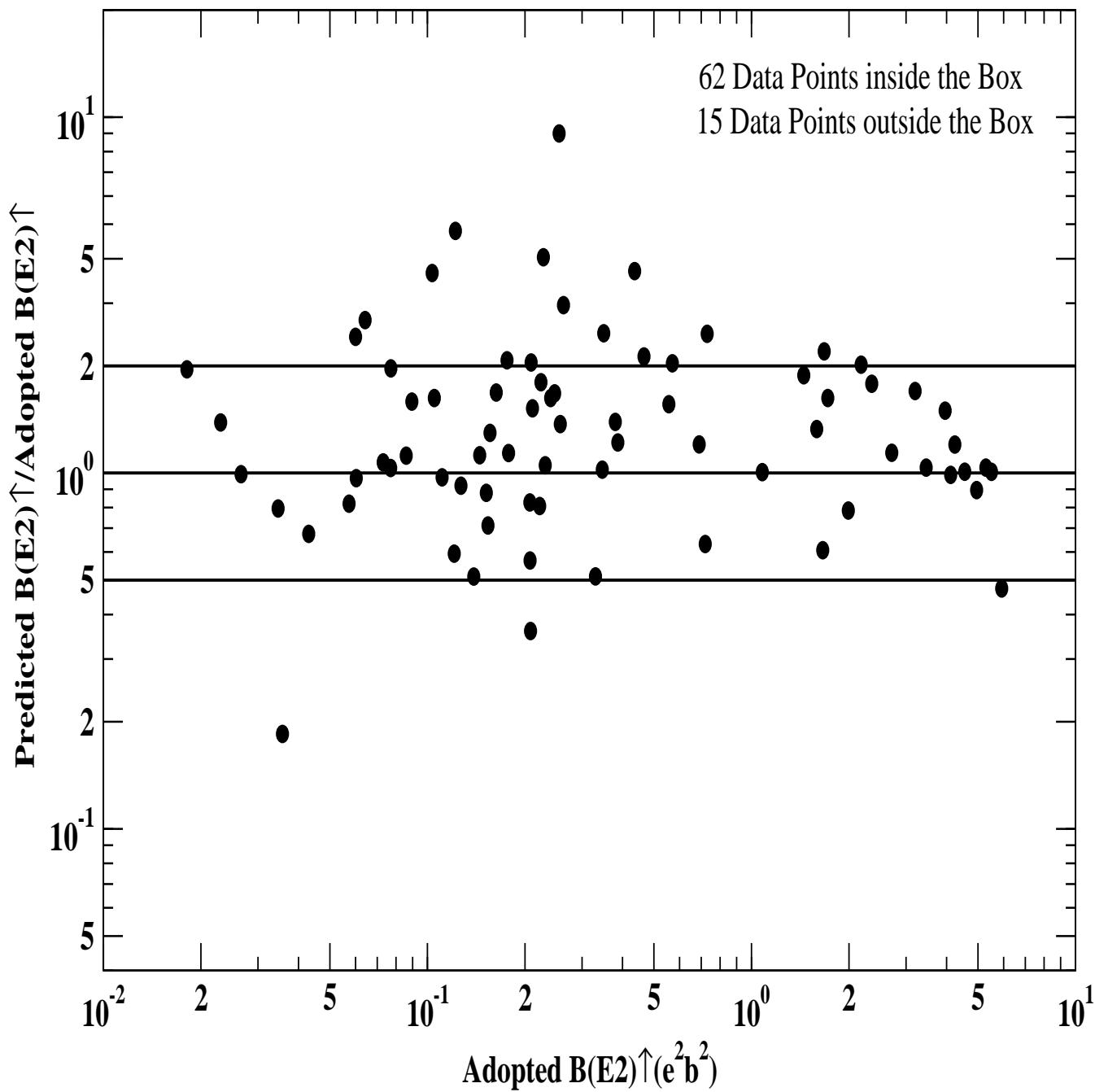}
\caption{Same as Fig. 13 but for  $B(E2)\uparrow$ compared with
with the latest adopted experimental  data\cite{pr2} (see text).
 The data points lying inside the box indicate the
degree of agreement within a factor of two. }
\label{b2l}
\end{figure}

\twocolumn
\pagestyle{fancy}
\setlength{\textwidth}{7.5in}
\setlength{\textheight}{9.35in}
\setlength{\topmargin}{-0.62in}
\setlength{\oddsidemargin}{-0.6in}
\setlength{\evensidemargin}{\oddsidemargin}

\setlength{\columnsep}{0.4in}
\renewcommand{\headrulewidth}{0.0pt}
\renewcommand{\footrulewidth}{0.0pt}\setlength{\headwidth}{\textwidth}
\setlength{\headheight}{1.65in}
\setlength{\headsep}{4pt}
\setlength{\topskip}{0pt}
\setlength{\footskip}{0pt}

\fancyhead[C]{
{
\qquad  }\\[28pt]
Table 1. Predicted   E2 \& B(E2)$\uparrow$ Values and the Corresponding 
   Calculated Deformation Parameters (See Text)\\
\rule{\textwidth}{0.9pt}\\[0pt]
\begin{tabular*}{1.00\textwidth}{p{0.1in}%
@{\extracolsep{\fill}}p{0.12in}p{0.15in}ccc!
{\mbox{\hspace{1pt}}}p{0.15in}p{0.15in}p{0.15in}ccccccc}
& A & N & $E2$ &$B(E2)\uparrow$ & $\beta_2~~$ &  $\beta_2/\beta_{2(sp)}$ &$~~~~~~~~ Q_{0}$ &  
& A & N & $E2$ &$B(E2)\uparrow$ & $\beta_2$ &  $\beta_2/\beta_{2(sp)}$ & $Q_{0}$   
 \end{tabular*}
 \rule{\textwidth}{0.9pt} }

\fancyfoot[C]{
 \vskip-47pt
 \rule{\textwidth}{0.4pt}\\
 \vskip10pt\thepage}


\renewcommand{\arraystretch}{1.}
\renewcommand{\baselinestretch}{1.25}

 \setlength{\tabcolsep}{4pt}
\tablefirsthead{\vspace{-21pt}\\}
 \tablehead{\vspace{-21pt}\\}
\tabletail{}
\tablelasttail{}

\normalsize
\newcolumntype{d}{D{.}{.}{-1}}

\begin{supertabular*}{0.99\columnwidth}{c@{\extracolsep{\fill}}crrddd@{}d}
\multicolumn{7}{l}{$Z =   8$ (O )} \\[2pt]
& 12&  4& 5.505& 0.010& 0.701& 3.536& 0.320 \\
& 14&  6& 4.529& 0.006& 0.497& 2.507& 0.252 \\
& 24& 16& 3.005& 0.014& 0.513& 2.591& 0.372 \\
\multicolumn{7}{l}{$Z =  10$ (Ne)} \\[2pt]

& 30& 20& 1.360& 0.038& 0.591& 3.729& 0.622 \\
& 32& 22& 0.948& 0.048& 0.632& 3.988& 0.695 \\
\multicolumn{7}{l}{$Z =  12$ (Mg)} \\[2pt]

& 18&  6& 1.910& 0.027& 0.579& 4.382& 0.520 \\
& 20&  8& 1.500& 0.030& 0.569& 4.306& 0.548 \\
& 34& 22& 0.678& 0.047& 0.501& 3.790& 0.687 \\
& 36& 24& 0.361& 0.016& 0.284& 2.149& 0.405 \\
& 38& 26& 0.396& 0.030& 0.371& 2.807& 0.548 \\
\multicolumn{7}{l}{$Z =  14$ (Si)} \\[2pt]

& 22&  8& 1.928& 0.021& 0.388& 3.427& 0.465 \\
& 24& 10& 1.673& 0.029& 0.428& 3.781& 0.544 \\
& 40& 26& 1.022& 0.029& 0.303& 2.672& 0.540 \\
& 42& 28& 1.894& 0.011& 0.184& 1.621& 0.339 \\
& 44& 30& 1.679& 0.027& 0.272& 2.406& 0.518 \\
\multicolumn{7}{l}{$Z =  16$ (S )} \\[2pt]

& 26& 10& 2.128& 0.029& 0.354& 3.573& 0.542 \\
& 28& 12& 2.228& 0.035& 0.371& 3.744& 0.597 \\
& 46& 30& 1.275& 0.040& 0.282& 2.848& 0.632 \\
& 48& 32& 1.234& 0.049& 0.304& 3.072& 0.701 \\
\multicolumn{7}{l}{$Z =  18$ (Ar)} \\[2pt]

& 30& 12& 2.057& 0.029& 0.283& 3.208& 0.535 \\
& 32& 14& 2.071& 0.026& 0.261& 2.958& 0.515 \\
& 48& 30& 1.510& 0.028& 0.203& 2.304& 0.526 \\
& 50& 32& 2.712& 0.037& 0.228& 2.590& 0.607 \\
\multicolumn{7}{l}{$Z =  20$ (Ca)} \\[2pt]

& 36& 16& 2.326& 0.011& 0.138& 1.741& 0.328 \\
\multicolumn{7}{l}{$Z =  20$ (Ca)} \\[2pt]
& 50& 30& 1.075& 0.037& 0.206& 2.594& 0.608 \\
& 52& 32& 1.079& 0.032& 0.186& 2.341& 0.564 \\
& 54& 34& 2.108& 0.366& 0.616& 7.766& 1.917 \\
\multicolumn{7}{l}{$Z =  22$ (Ti)} \\[2pt]

& 38& 16& 2.655& 0.092& 0.355& 4.924& 0.962 \\
& 40& 18& 1.862& 0.080& 0.320& 4.440& 0.897 \\
& 52& 30& 1.053& 0.058& 0.229& 3.180& 0.766 \\
& 54& 32& 1.386& 0.007& 0.075& 1.043& 0.258 \\
& 56& 34& 1.182& 0.145& 0.344& 4.770& 1.207 \\
& 58& 36& 0.843& 0.155& 0.347& 4.817& 1.247 \\
& 60& 38& 0.552& 0.353& 0.513& 7.116& 1.885 \\
& 62& 40& 0.491& 0.066& 0.216& 3.000& 0.812 \\
\multicolumn{7}{l}{$Z =  24$ (Cr)} \\[2pt]
& 42& 18& 0.999& 0.116& 0.341& 5.166& 1.078 \\
& 44& 20& 1.364& 0.115& 0.330& 4.993& 1.075 \\
& 60& 36& 0.818& 0.108& 0.260& 3.933& 1.042 \\
& 62& 38& 0.490& 0.274& 0.405& 6.134& 1.660 \\
& 64& 40& 0.349& 0.202& 0.341& 5.158& 1.426 \\
& 66& 42& 0.366 \\
\multicolumn{7}{l}{$Z =  26$ (Fe)} \\[2pt]

& 46& 20& 1.422& 0.080& 0.246& 4.032& 0.895 \\
& 46& 20& 1.422& 0.080& 0.246& 4.032& 0.895 \\
& 48& 22& 0.905& 0.060& 0.208& 3.414& 0.779 \\
& 50& 24& 0.842& 0.179& 0.349& 5.722& 1.342 \\
& 52& 26& 0.895& 0.092& 0.243& 3.993& 0.961 \\
& 66& 40& 1.200& 0.134& 0.250& 4.108& 1.159 \\
& 68& 42& 0.413& 0.203& 0.302& 4.956& 1.427 \\
& 70& 44& 0.421 \\
& 72& 46& 0.833 \\
& 74& 48& 0.866 \\
\\
\\
\multicolumn{7}{l}{$Z =  28$ (Ni)} \\[2pt]

& 50& 22& 1.217 \\
& 52& 24& 1.308& 0.148& 0.287& 5.070& 1.221 \\
& 70& 42& 1.397& 0.096& 0.190& 3.351& 0.983 \\
& 72& 44& 1.020& 0.225& 0.285& 5.027& 1.503 \\
& 74& 46& 0.942& 0.173& 0.245& 4.326& 1.318 \\
& 78& 50& 1.924 \\
\multicolumn{7}{l}{$Z =  30$ (Zn)} \\[2pt]

& 54& 24& 1.329& 0.096& 0.210& 3.971& 0.980 \\
& 56& 26& 0.277& 0.080& 0.187& 3.544& 0.896 \\
& 58& 28& 2.070& 0.026& 0.105& 1.994& 0.516 \\
& 60& 30& 1.275& 0.095& 0.195& 3.690& 0.977 \\
& 60& 30& 1.275& 0.095& 0.195& 3.690& 0.977 \\
& 76& 46& 0.745& 0.163& 0.218& 4.125& 1.279 \\
& 78& 48& 0.800& 0.080& 0.150& 2.834& 0.894 \\
& 80& 50& 1.573& 0.078& 0.146& 2.764& 0.887 \\
& 82& 52& 1.638& 0.085& 0.150& 2.838& 0.925 \\
\multicolumn{7}{l}{$Z =  32$ (Ge)} \\[2pt]

& 62& 30& 0.956& 0.048& 0.127& 2.572& 0.696 \\
& 64& 32& 0.923& 0.075& 0.155& 3.137& 0.867 \\
& 78& 46& 0.585& 0.179& 0.211& 4.256& 1.343 \\
& 80& 48& 0.828& 0.071& 0.131& 2.635& 0.845 \\
& 82& 50& 1.414& 0.072& 0.129& 2.605& 0.850 \\
& 84& 52& 1.334& 0.097& 0.148& 2.980& 0.987 \\
\multicolumn{7}{l}{$Z =  34$ (Se)} \\[2pt]

& 64& 30& 0.856& 0.263& 0.274& 5.878& 1.625 \\
& 66& 32& 0.802& 0.289& 0.282& 6.043& 1.705 \\
& 68& 34& 0.889& 0.321& 0.291& 6.236& 1.795 \\
& 84& 50& 1.234& 0.171& 0.184& 3.950& 1.309 \\
& 86& 52& 1.182& 0.187& 0.190& 4.070& 1.370 \\
& 88& 54& 0.625& 0.054& 0.101& 2.161& 0.739 \\
& 90& 56& 0.823& 0.124& 0.150& 3.219& 1.117 \\
& 92& 58& 0.461 \\
\multicolumn{7}{l}{$Z =  36$ (Kr)} \\[2pt]

& 70& 34& 0.624& 0.913& 0.455&10.323& 3.030 \\

& 72& 36& 0.567& 0.991& 0.465&10.552& 3.156 \\
& 88& 52& 1.138& 0.142& 0.154& 3.496& 1.195 \\
& 90& 54& 0.762& 0.109& 0.133& 3.022& 1.049 \\
& 92& 56& 0.734& 0.118& 0.136& 3.089& 1.088 \\
& 94& 58& 0.746& 0.414& 0.251& 5.708& 2.039 \\
& 96& 60& 0.595& 1.609& 0.489&11.103& 4.022 \\
& 98& 62& 0.099 \\
&100& 64& 0.113 \\
\multicolumn{7}{l}{$Z =  38$ (Sr)} \\[2pt]

& 74& 36& 0.527& 1.236& 0.483&11.573& 3.525 \\
& 76& 38& 0.321& 1.086& 0.445&10.660& 3.305 \\
&102& 64& 0.118& 1.319& 0.403& 9.652& 3.641 \\
&104& 66& 0.121 \\
\multicolumn{7}{l}{$Z =  40$ (Zr)} \\[2pt]

& 76& 36& 0.538 \\
& 78& 38& 0.272& 1.051& 0.408&10.304& 3.251 \\
& 80& 40& 0.294& 1.042& 0.400&10.089& 3.237 \\
& 98& 58& 1.088& 0.463& 0.233& 5.873& 2.157 \\
&104& 64& 0.136& 1.561& 0.411&10.367& 3.961 \\
&106& 66& 0.171& 1.980& 0.457&11.526& 4.461 \\
&108& 68& 0.175& 1.961& 0.449&11.332& 4.441 \\
\multicolumn{7}{l}{$Z =  42$ (Mo)} \\[2pt]

& 82& 40& 0.323& 1.575& 0.460&12.200& 3.979 \\
& 84& 42& 0.427& 1.451& 0.435&11.524& 3.819 \\
& 86& 44& 0.550& 0.759& 0.310& 8.206& 2.762 \\
& 88& 46& 0.688& 0.066& 0.090& 2.386& 0.816 \\
& 90& 48& 0.942& 0.318& 0.194& 5.149& 1.787 \\
&110& 68& 0.194& 1.650& 0.388&10.268& 4.073 \\
&112& 70& 0.173& 1.769& 0.396&10.503& 4.217 \\
\multicolumn{7}{l}{$Z =  44$ (Ru)} \\[2pt]

& 86& 42& 0.490 \\

\multicolumn{7}{l}{$Z =  44$ (Ru)} \\[2pt]

& 92& 48& 0.906& 0.380& 0.200& 5.547& 1.953 \\
& 94& 50& 1.390& 0.141& 0.120& 3.333& 1.191 \\
&114& 70& 0.260& 1.337& 0.325& 9.024& 3.666 \\
&116& 72& 0.421& 1.058& 0.286& 7.935& 3.261 \\
&118& 74& 0.535 \\
\multicolumn{7}{l}{$Z =  46$ (Pd)} \\[2pt]

& 90& 44& 0.522 \\
& 92& 46& 0.685 \\
& 96& 50& 1.315& 0.011& 0.032& 0.926& 0.336 \\
& 98& 52& 0.957& 0.096& 0.092& 2.672& 0.981 \\
&100& 54& 0.676& 0.354& 0.175& 5.069& 1.887 \\
&118& 72& 0.383& 0.500& 0.186& 5.393& 2.242 \\
&120& 74& 0.476& 0.710& 0.219& 6.355& 2.672 \\
&122& 76& 0.657 \\
&124& 78& 0.688 \\
&126& 80& 0.875 \\
&128& 82& 0.937 \\
\\
\multicolumn{7}{l}{$Z =  48$ (Cd)} \\[2pt]

& 96& 48& 0.770 \\
&100& 52& 1.055& 0.169& 0.116& 3.501& 1.303 \\
&102& 54& 0.820& 0.352& 0.165& 4.990& 1.882 \\
&124& 76& 0.719& 0.864& 0.227& 6.860& 2.948 \\
&126& 78& 0.731& 0.779& 0.213& 6.445& 2.799 \\
&128& 80& 0.804 \\
&130& 82& 0.720 \\
\multicolumn{7}{l}{$Z =  50$ (Sn)} \\[2pt]

&100& 50& 1.862 \\
&104& 54& 1.252& 0.365& 0.159& 5.011& 1.915 \\
&106& 56& 1.071& 0.427& 0.170& 5.354& 2.072 \\
&108& 58& 1.108& 0.403& 0.163& 5.136& 2.013 \\
&110& 60& 1.115& 0.243& 0.125& 3.939& 1.563 \\
\multicolumn{7}{l}{$Z =  50$ (Sn)} \\[2pt]
&126& 76& 1.046& 0.117& 0.079& 2.496& 1.084 \\
&128& 78& 1.101& 0.152& 0.089& 2.814& 1.235 \\
&130& 80& 1.276& 0.032& 0.040& 1.277& 0.566 \\
&136& 86& 0.485 \\
\multicolumn{7}{l}{$Z =  52$ (Te)} \\[2pt]
&106& 54& 0.715 \\
&108& 56& 0.701& 0.471& 0.169& 5.555& 2.177 \\
&110& 58& 0.738& 0.605& 0.189& 6.215& 2.465 \\
&112& 60& 0.702& 0.835& 0.220& 7.215& 2.897 \\
&114& 62& 0.691& 0.868& 0.222& 7.272& 2.954 \\
&116& 64& 0.688& 0.931& 0.227& 7.445& 3.060 \\
&118& 66& 0.691& 1.159& 0.250& 8.211& 3.413 \\

&132& 80& 1.039& 0.171& 0.089& 2.928& 1.312 \\
&134& 82& 1.109& 0.377& 0.131& 4.303& 1.947 \\
&136& 84& 0.954& 0.584& 0.162& 5.304& 2.424 \\
&138& 86& 0.328& 0.633& 0.167& 5.466& 2.522 \\
&140& 88& 0.312& 1.229& 0.230& 7.545& 3.515 \\
&142& 90& 0.186 \\
\multicolumn{7}{l}{$Z =  54$ (Xe)} \\[2pt]

&110& 56& 0.443& 0.346& 0.138& 4.701& 1.865 \\
&112& 58& 0.434& 0.669& 0.190& 6.458& 2.593 \\
&138& 84& 0.798& 0.529& 0.147& 4.997& 2.306 \\
&142& 88& 0.313& 0.830& 0.180& 6.142& 2.889 \\
&144& 90& 0.246& 1.797& 0.263& 8.953& 4.250 \\
&146& 92& 0.186 \\
\multicolumn{7}{l}{$Z =  56$ (Ba)} \\[2pt]

&114& 58& 0.501& 0.373& 0.135& 4.768& 1.937 \\
&116& 60& 0.268& 0.653& 0.177& 6.236& 2.563 \\
&118& 62& 0.257& 1.037& 0.220& 7.766& 3.228 \\
&120& 64& 0.239& 2.351& 0.327&11.565& 4.862 \\
&148& 92& 0.199& 3.160& 0.330&11.658& 5.636 \\
&150& 94& 0.113& 2.524& 0.292&10.326& 5.037 \\
\multicolumn{7}{l}{$Z =  58$ (Ce)} \\[2pt]

&120& 62& 0.213& 3.032& 0.359&13.132& 5.521 \\
&122& 64& 0.132& 4.370& 0.426&15.594& 6.628 \\
&142& 84& 0.954& 0.394& 0.116& 4.234& 1.991 \\
&144& 86& 0.415& 0.863& 0.170& 6.205& 2.945 \\
&152& 94& 0.090& 2.804& 0.295&10.789& 5.310 \\
&154& 96& 0.074& 2.011& 0.248& 9.056& 4.496 \\
\multicolumn{7}{l}{$Z =  60$ (Nd)} \\[2pt]

&124& 64& 0.099& 5.634& 0.463&17.516& 7.526 \\
&126& 66& 0.107& 4.988& 0.431&16.306& 7.082 \\
&128& 68& 0.127& 4.392& 0.400&15.141& 6.645 \\
&136& 76& 0.381& 1.008& 0.184& 6.965& 3.183 \\
&138& 78& 0.544& 1.322& 0.209& 7.901& 3.646 \\
&140& 80& 0.991& 0.455& 0.121& 4.590& 2.138 \\
&154& 94& 0.070& 4.187& 0.345&13.069& 6.488 \\
&156& 96& 0.067& 5.011& 0.374&14.174& 7.098 \\
&158& 98& 0.065 \\
\multicolumn{7}{l}{$Z =  62$ (Sm)} \\[2pt]

&128& 66& 0.108 \\
&130& 68& 0.107& 5.087& 0.412&16.127& 7.151 \\
&132& 70& 0.119& 4.811& 0.397&15.525& 6.955 \\
&140& 78& 0.521& 0.862& 0.162& 6.319& 2.944 \\
&146& 84& 1.008& 0.390& 0.106& 4.132& 1.980 \\
&156& 94& 0.075& 4.844& 0.356&13.936& 6.978 \\
&158& 96& 0.072& 5.431& 0.374&14.631& 7.389 \\
&160& 98& 0.067& 5.745& 0.382&14.923& 7.600 \\
&162&100& 0.072& 5.594& 0.373&14.604& 7.499 \\
\\
\multicolumn{7}{l}{$Z =  64$ (Gd)} \\[2pt]

&132& 68& 0.116 \\
&134& 70& 0.099 \\
&136& 72& 0.123 \\
&138& 74& 0.229& 4.394& 0.357&14.403& 6.646 \\

\multicolumn{7}{l}{$Z =  64$ (Gd)} \\[2pt]

&140& 76& 0.331& 3.074& 0.296&11.931& 5.559 \\
&142& 78& 0.506& 0.241& 0.082& 3.313& 1.558 \\
&144& 80& 1.125& 0.250& 0.083& 3.341& 1.586 \\
&146& 82& 1.618& 0.101& 0.052& 2.107& 1.010 \\
&148& 84& 1.228& 0.920& 0.156& 6.292& 3.042 \\
&150& 86& 0.583& 0.997& 0.161& 6.490& 3.166 \\
&162& 98& 0.069& 5.547& 0.360&14.543& 7.468 \\
&164&100& 0.075& 5.479& 0.355&14.335& 7.421 \\
&166&102& 0.073& 5.728& 0.360&14.539& 7.588 \\
&168&104& 0.065 \\
\multicolumn{7}{l}{$Z =  66$ (Dy)} \\[2pt]

&140& 74& 0.200 \\
&150& 84& 1.074& 0.241& 0.077& 3.193& 1.558 \\
&166&100& 0.073& 5.560& 0.344&14.325& 7.476 \\
&168&102& 0.075& 5.694& 0.345&14.381& 7.566 \\
&170&104& 0.072& 5.542& 0.338&14.076& 7.464 \\
\multicolumn{7}{l}{$Z =  68$ (Er)} \\[2pt]

&142& 74& 0.334 \\
&146& 78& 0.431 \\
&152& 84& 1.029& 0.564& 0.113& 4.839& 2.382 \\
&154& 86& 0.581& 0.512& 0.106& 4.568& 2.268 \\
&172&104& 0.076& 5.705& 0.330&14.170& 7.573 \\
&174&106& 0.084& 4.179& 0.281&12.036& 6.482 \\
&176&108& 0.085& 3.987& 0.272&11.667& 6.331 \\
\multicolumn{7}{l}{$Z =  70$ (Yb)} \\[2pt]

&150& 80& 1.454 \\
&154& 84& 0.865& 3.184& 0.258&11.396& 5.657 \\
&156& 86& 0.524& 2.010& 0.203& 8.978& 4.496 \\
&178&108& 0.084& 5.016& 0.294&12.987& 7.101 \\
&180&110& 0.091& 4.176& 0.266&11.763& 6.480 \\
&182&112& 0.119& 1.219& 0.143& 6.307& 3.500 \\
\\
\multicolumn{7}{l}{$Z =  72$ (Hf)} \\[2pt]

&152& 80& 1.460 \\
&158& 86& 0.633& 1.120& 0.146& 6.645& 3.356 \\
&160& 88& 0.402& 0.601& 0.106& 4.827& 2.458 \\
&182&110& 0.108& 4.062& 0.254&11.515& 6.390 \\
&184&112& 0.113& 2.982& 0.216& 9.794& 5.475 \\
&186&114& 0.135& 2.537& 0.197& 8.970& 5.050 \\
&188&116& 0.297 \\
\multicolumn{7}{l}{$Z =  74$ (W )} \\[2pt]

&156& 82& 2.301 \\
&158& 84& 1.327 \\
&160& 86& 0.542 \\
&164& 90& 0.339& 0.400& 0.083& 3.873& 2.005 \\
&166& 92& 0.256& 1.714& 0.170& 7.953& 4.151 \\
&176&102& 0.110& 4.433& 0.264&12.301& 6.676 \\
&178&104& 0.105& 4.591& 0.266&12.425& 6.794 \\
&188&114& 0.176& 3.089& 0.211& 9.827& 5.573 \\
&190&116& 0.173& 1.663& 0.153& 7.160& 4.089 \\
&192&118& 0.220& 1.206& 0.130& 6.055& 3.482 \\
&194&120& 0.421 \\
\multicolumn{7}{l}{$Z =  76$ (Os)} \\[2pt]

&162& 86& 0.620 \\
&168& 92& 0.356& 1.393& 0.148& 7.114& 3.742 \\
&170& 94& 0.290& 3.025& 0.217&10.399& 5.514 \\
&176&100& 0.113& 5.438& 0.284&13.624& 7.394 \\
&178&102& 0.128& 4.071& 0.244&11.700& 6.397 \\
&194&118& 0.279& 1.686& 0.148& 7.110& 4.117 \\
&196&120& 0.305& 0.599& 0.088& 4.207& 2.453 \\
&198&122& 0.397& 0.243& 0.056& 2.663& 1.563 \\
\multicolumn{7}{l}{$Z =  78$ (Pt)} \\[2pt]

&166& 88& 0.699 \\
&172& 94& 0.434& 1.382& 0.142& 6.975& 3.727 \\
&174& 96& 0.359& 1.229& 0.133& 6.527& 3.515 \\
\multicolumn{7}{l}{$Z =  78$ (Pt)} \\[2pt]

&176& 98& 0.287& 2.580& 0.191& 9.385& 5.093 \\
&178&100& 0.182& 5.098& 0.266&13.093& 7.159 \\
&180&102& 0.187& 4.810& 0.257&12.623& 6.954 \\
&182&104& 0.183& 3.583& 0.220&10.816& 6.002 \\
&200&122& 0.489& 0.737& 0.094& 4.607& 2.723 \\
&202&124& 0.459& 0.130& 0.039& 1.919& 1.141 \\
&204&126& 2.037& 0.457& 0.073& 3.581& 2.144 \\
\multicolumn{7}{l}{$Z =  80$ (Hg)} \\[2pt]

&170& 90& 0.681 \\
&172& 92& 0.583 \\
&174& 94& 0.664 \\
&180&100& 0.418& 2.729& 0.188& 9.508& 5.238 \\
&182&102& 0.338& 3.684& 0.217&10.967& 6.086 \\
&188&108& 0.414& 2.793& 0.185& 9.345& 5.299 \\
&190&110& 0.464& 4.085& 0.222&11.221& 6.408 \\

\multicolumn{7}{l}{$Z =  80$ (Hg)} \\[2pt]

&192&112& 0.498& 1.623& 0.139& 7.023& 4.039 \\
&194&114& 0.506& 1.521& 0.134& 6.753& 3.910 \\
&206&126& 0.638& 0.622& 0.082& 4.150& 2.501 \\
&208&128& 0.788& 0.418& 0.067& 3.381& 2.051 \\
&210&130& 0.793& 0.082& 0.029& 1.487& 0.908 \\
&212&132& 0.866 \\
\multicolumn{7}{l}{$Z =  82$ (Pb)} \\[2pt]

&178& 96& 0.695 \\
&180& 98& 0.984 \\
&186&104& 0.614& 2.933& 0.186& 9.645& 5.430 \\
&188&106& 0.665& 2.293& 0.164& 8.467& 4.801 \\
&190&108& 0.693& 1.120& 0.114& 5.876& 3.356 \\
&192&110& 0.772& 3.963& 0.212&10.976& 6.312 \\
&194&112& 0.881& 1.805& 0.142& 7.357& 4.260 \\
&196&114& 0.947& 1.829& 0.142& 7.355& 4.288 \\
&198&116& 0.947& 1.411& 0.124& 6.415& 3.766 \\
\multicolumn{7}{l}{$Z =  82$ (Pb)} \\[2pt]
&200&118& 0.919& 1.217& 0.114& 5.918& 3.498 \\
&202&120& 0.833& 0.618& 0.081& 4.190& 2.492 \\
&212&130& 0.671& 0.064& 0.025& 1.302& 0.800 \\
&216&134& 0.732 \\
\multicolumn{7}{l}{$Z =  84$ (Po)} \\[2pt]

&184&100& 1.138 \\
&186&102& 0.767 \\
&188&104& 0.601 \\
&190&106& 0.677 \\
&196&112& 0.557& 2.112& 0.149& 7.904& 4.608 \\
&198&114& 0.636& 2.184& 0.151& 7.983& 4.686 \\
&200&116& 0.716& 1.719& 0.133& 7.035& 4.158 \\
&202&118& 0.741& 2.108& 0.146& 7.739& 4.604 \\
&204&120& 0.744& 0.810& 0.090& 4.766& 2.854 \\
&206&122& 0.763& 0.258& 0.050& 2.673& 1.611 \\
&208&124& 0.775& 0.196& 0.044& 2.315& 1.404 \\
&212&128& 0.820& 0.033& 0.018& 0.932& 0.572 \\
&220&136& 0.392 \\
\multicolumn{7}{l}{$Z =  86$ (Rn)} \\[2pt]

&194&108& 0.196 \\
&196&110& 0.187 \\
&224&138& 0.147& 4.116& 0.186&10.093& 6.433 \\
&226&140& 0.056& 5.361& 0.211&11.451& 7.341 \\
&228&142& 0.046 \\
\multicolumn{7}{l}{$Z =  88$ (Ra)} \\[2pt]

&200&112& 0.225 \\
&202&114& 0.349 \\
&204&116& 0.439 \\
&220&132& 0.279& 5.062& 0.204&11.329& 7.134 \\
&230&142& 0.057& 7.909& 0.248&13.746& 8.917 \\
&232&144& 0.056& 6.472& 0.223&12.364& 8.066 \\
&234&146& 0.053& 7.747& 0.242&13.449& 8.825 \\
\multicolumn{7}{l}{$Z =  88$ (Ra)} \\[2pt]

&236&148& 0.037& 9.642& 0.269&14.920& 9.845 \\
\multicolumn{7}{l}{$Z =  90$ (Th)} \\[2pt]

&208&118& 0.423 \\
&210&120& 0.419 \\
&212&122& 0.581 \\
&214&124& 1.065 \\
&218&128& 0.886& 0.551& 0.066& 3.759& 2.353 \\
&220&130& 0.488& 0.782& 0.078& 4.454& 2.805 \\
&224&134& 0.112& 5.927& 0.213&12.111& 7.719 \\
&236&146& 0.048& 9.222& 0.257&14.591& 9.629 \\
&238&148& 0.044&10.482& 0.272&15.469&10.265 \\
\multicolumn{7}{l}{$Z =  92$ (U )} \\[2pt]

&218&126& 2.615 \\
&220&128& 1.155 \\
&222&130& 0.656 \\
&224&132& 0.190& 4.238& 0.176&10.242& 6.527 \\
&226&134& 0.094& 7.505& 0.233&13.548& 8.686 \\
&228&136& 0.073& 9.002& 0.254&14.752& 9.513 \\
&240&148& 0.046&12.691& 0.292&16.927&11.295 \\
&242&150& 0.047&13.836& 0.303&17.576&11.794 \\
&244&152& 0.040&14.057& 0.304&17.619&11.888 \\
&246&154& 0.040&15.872& 0.321&18.620&12.632 \\
\multicolumn{7}{l}{$Z =  94$ (Pu)} \\[2pt]

&228&134& 0.114 \\
&230&136& 0.071& 8.664& 0.243&14.388& 9.333 \\
&232&138& 0.059& 9.461& 0.252&14.948& 9.753 \\
&234&140& 0.050& 9.213& 0.247&14.667& 9.624 \\
&236&142& 0.042&11.257& 0.272&16.121&10.638 \\
&246&152& 0.046&13.915& 0.294&17.434&11.827 \\
&248&154& 0.043&15.366& 0.307&18.222&12.429 \\
\multicolumn{7}{l}{$Z =  96$ (Cm)} \\[2pt]

&236&140& 0.028&12.032& 0.275&16.667&10.998 \\
\multicolumn{7}{l}{$Z =  96$ (Cm)} \\[2pt]

&238&142& 0.040&15.228& 0.308&18.645&12.373 \\
&242&146& 0.041&13.970& 0.292&17.661&11.851 \\
&250&154& 0.047&16.004& 0.305&18.498&12.684 \\
&252&156& 0.042&15.778& 0.302&18.269&12.595 \\
\multicolumn{7}{l}{$Z =  98$ (Cf)} \\[2pt]

&236&138& 0.020& 7.351& 0.211&13.028& 8.597 \\
&238&140& 0.023&12.734& 0.276&17.050&11.314 \\
&240&142& 0.031&16.918& 0.316&19.543&13.041 \\
&242&144& 0.035&15.093& 0.297&18.357&12.318 \\
&244&146& 0.042&13.743& 0.282&17.421&11.754 \\
&246&148& 0.042&14.559& 0.288&17.834&12.098 \\
&248&150& 0.043&15.807& 0.299&18.482&12.606 \\
&254&156& 0.049 \\
\multicolumn{7}{l}{$Z = 100$ (Fm)} \\[2pt]

&246&146& 0.041&12.806& 0.265&16.725&11.347 \\
&248&148& 0.045&12.973& 0.265&16.743&11.420 \\
&250&150& 0.044&15.834& 0.292&18.399&12.617 \\
&252&152& 0.042&15.241& 0.285&17.955&12.378 \\
 \end{supertabular*}
\end{document}